\documentclass[prc,twocolumn,superscriptaddress,showpacs,floatfix]{revtex4}

\usepackage{graphicx}

\newcommand{\be}{\begin{equation}}
\newcommand{\ee}{  \end{equation}}
\newcommand{\ba}{\begin{eqnarray}}
\newcommand{\ea}{  \end{eqnarray}}
\newcommand{\ve}{\varepsilon}

\begin{document}

\title{The Two--Body Random Ensemble in Nuclei}

\author{T.~Papenbrock}
\affiliation{Department of Physics and Astronomy, University of Tennessee,
Knoxville, TN~37996, USA}
\affiliation{Physics Division,
Oak Ridge National Laboratory, Oak Ridge, TN 37831, USA}
\author{H.~A.~Weidenm\"uller}
\affiliation{Max-Planck Institut f\"ur Kernphysik,
D-69029 Heidelberg, Germany}
\date{\today}

\begin{abstract}
Combining analytical and numerical methods, we investigate properties
of the two--body random ensemble (TBRE). We compare the TBRE with the
Gaussian orthogonal ensemble of random matrices. Using the geometric
properties of the nuclear shell model, we discuss the information
content of nuclear spectra, and gain insight in the difficulties
encountered when fitting the effective interaction. 
We exhibit the
existence of correlations between spectral widths pertaining to
different quantum numbers. Using these results, we deduce the
preponderance of spin--zero ground states in the TBRE. We demonstrate
the existence of correlations between spectra with different quantum
numbers and/or in different nuclei.
\end{abstract}

\maketitle

\section{Introduction}

The spectral fluctuation properties of complex nuclear spectra often
agree with predictions of random--matrix theory, more precisely: With
those of the Gaussian orthogonal ensemble of random matrices (GOE).
This is true for the resonances observed at neutron threshold and at
the Coulomb barrier for protons~\cite{HPB} but applies in a number of
cases likewise to levels at lower excitation energies, see
Ref.~\cite{AHSW} and references therein.

The GOE is an ensemble of matrices where every state in Hilbert space
interacts with every other state (if both carry the same conserved
quantum numbers). Indeed, the probability density ${\cal P}(H_{\rm GOE})$
of the real and symmetric Hamiltonian matrices $H_{\rm GOE}$ of the GOE
has the form
\be
{\cal P}(H_{\rm GOE}) = {\cal N} \exp \big\{- \frac{N}{\lambda^2}
{\rm Trace} [ H_{\rm GOE}^2 ] \big\}
\label{GOE}
\ee
where $N$ denotes the dimension of the matrices and $\lambda$ a
parameter which determines the average level density while ${\cal N}$
is a normalization factor. Eq.~(\ref{GOE}) shows that the matrix
elements connecting any pair of states are uncorrelated
Gaussian--distributed random variables.

This basic tenet of the GOE is not in keeping with the shell model,
the fundamental dynamical model of nuclear physics~\cite{MJ}. Indeed,
that model is basically a single--particle model with a residual
interaction. The interaction is dominated by two--body forces. In a
representation where the many--body states are Slater determinants of
single--particle states, a two--body interaction will have non--zero
matrix elements only between those Slater determinants which differ by
at most two units in the occupation numbers of the single--particle
states. Out of the total number of such determinants, this is a small
fraction. Put differently: In an arbitrary basis for the many--body
states, the number of independent matrix elements of the two--body
interaction is very much smaller than that of the GOE. This fact is
changed only quantitatively but not qualitatively when we allow for
a three--body residual interaction.

Already in the 1970's, this fact has led to the question: Are the
predictions of the GOE for spectral fluctuation properties in keeping
with the results of shell--model calculations with a residual two--body
interaction? The answer, based on numerical calculations, has been
affirmative~\cite{FW,BF}, and numerous more recent calculations have
confirmed it (see the review~\cite{ZE} and references therein). The
calculations were based upon a random--matrix ensemble (the two--body
random ensemble (TBRE)) which differs from the GOE and accounts for the
specific properties of the nuclear shell model: The existence of a
residual two--body interaction which conserves total spin, parity, and
isospin.

In spite of the agreement between GOE predictions and numerical
results for the TBRE, open questions remain. In fact, analytical
studies of the TBRE are virtually non--existent. Treating the TBRE
analytically poses a severe challenge. We use a combination of
numerical and analytical techniques to display some of its
properties. We focus on the underlying geometric structure of the TBRE
and its implications for correlations between spectra of different
quantum numbers (like nucleon number and/or spin). The results we are
obtain are interesting also from a practical point of view, as we gain
insight into the inner workings of the shell model and are able to
separate geometric and dynamical properties. Together with the work
reported in Refs.~\cite{PW,PW1}, the present paper may be seen as a
step towards filling the gap in understanding the TBRE.

In treating the TBRE, we neglect the influence of possible three--body
forces. It will become clear below that this omission affects our
results in a quantitative but not in a qualitative fashion. Spectral
fluctuation properties of the GOE type emerge only when there are at
least three nucleons in a major shell. This is why we confine ourselves
to this case.

This paper is organized as follows.  The two--body random ensemble is
defined and compared with the GOE in Section~\ref{def}. Some of basic
properties of the TBRE are displayed in Section~\ref{prop}. This main
section contains results about the information content of nuclear
spectra, the preponderance of spin-0 ground states, and correlations
between spectra with different quantum numbers. Section~\ref{sum}
contains a brief summary. Technical details are relegated to several
Appendices.

\section{Definition of the TBRE}
\label{def}

The two--body random ensemble (TBRE) is defined within the framework
of the spherical nuclear shell model~\cite{MJ}. In that model, nucleons
move independently in a central potential with a strong spin--orbit
force. We consider one of the major shells of that model. Our numerical
examples are calculated for the $sd$--shell, with single--particle
states labelled $s_{1/2}, d_{3/2}$, and $d_{5/2}$ and single--particle
energies $\ve_{1/2}, \ve_{3/2}$, and $\ve_{5/2}$. Our general
considerations apply likewise, however, to other major shells in
heavier nuclei. There, however, the number of many--particle states
becomes forbiddingly large for numerical work. We sometimes also
consider a single $j$--shell with half--integer single--particle total
spin $j$. Although not realistic for nuclei, this case is sufficiently
simple to yield useful insights.

Putting several nucleons into a major shell, we construct a basis of
orthonormal antisymmetrized many--body states of fixed total spin $J$,
parity $P$, and isospin $T$. These states are labelled $| {\bf J} \mu
\rangle$, with ${\bf J}$ standing for the three quantum numbers $J, P,
T$ and with $\mu = 1, \ldots, d({\bf J})$ a running index with range
given by the dimension $d({\bf J})$ of Hilbert space ${\cal H}({\bf
J})$. We focus attention on a fixed but arbitrary $z$--projection $M$
of total spin $J$ so that $d({\bf J})$ is the actual number of states
not counting their degeneracy regarding $M$. In the middle of the
$sd$--shell and for low values of $J$, $d({\bf J})$ is typically of
order $10^3$ and much larger for heavier nuclei (other major shells).
The actual construction used for the basis of states $| {\bf J} \mu
\rangle$ is immaterial for what follows: The bases resulting from
different modes of construction are connected by a unitary transformation.
In the $sd$--shell, all single--particle states have positive parity, and
it is not neccessary to carry the quantum number $P$. Likewise, we often
consider only $sd$--shell states with isospin $T = 0$. In that case, it
suffices to label the many--body states by the total spin $J$ only. 

The number of nucleons in the major shell is denoted by $m$. Sometimes,
we will consider simultaneously several nuclei with different values of
$m$. In this case, we will denote, f.i., the dimension of Hilbert space
by $d({\bf J}, m)$, and similarly for other quantities.

The many--body states $| {\bf J} \mu \rangle$ are eigenstates of the
single--particle shell--model Hamiltonian with a very high degree of
degeneracy. The degeneracy is lifted when we take account of the
residual interaction of the shell model. We assume that this interaction
mixes states only within the same major shell. That assumption is rather
unrealistic because intruder states from higher shells occur even at low
excitation energies, and mixing with higher shells is bound to play a
major role at the upper end of the spectrum. The approximation is
perfectly adequate, however, for the purposes of this paper. For the
same reason we confine ourselves to a two--body residual interaction
(although there is evidence that three--body forces do play a role in
realistic attempts to model the spectrum), and we omit the Coulomb
interaction between protons. Finally, to focus attention on the role of
the residual interaction, we assume that the single--particle energies
within a major shell are all degenerate. For the $sd$--shell, this means
that we put $\ve_{1/2} = \ve_{3/2} = \ve_{5/2} = 0$. Then, the full
Hamiltonian $H$ of the shell model is determined entirely by the residual
interaction. In the sequel we consider the matrix elements $H_{\mu
\nu}({\bf J})$ of that Hamiltonian taken with respect to the basis of
states $| {\bf J} \mu \rangle$.

Within a major shell, the residual two--body interaction $V_2$
possesses a finite number of two--body matrix elements. These have the
following form. Let ${\bf j}_i$, $i = 1,2,3,4$ denote 4 (equal or
different) values of total single--particle spin, parity, and isospin
$1/2$. Coupling ${\bf j}_1$ and ${\bf j}_2$ (${\bf j}_3$ and ${\bf
j}_4$) to total two--body spin $s_1$ ($s_2$, repectively) and isospin
$t_1$ ($t_2$, respectively), denoting the parity of the resulting wave
functions by $\pi_1$ ($\pi_2$, respectively), and introducing the
notation ${\bf s}$ for the quantum numbers $s, \pi, t$, the reduced
two--body matrix elements of $V_2$ within a major shell have the form
$\langle {\bf j}_3 {\bf j}_4 {\bf s} || V_2 || {\bf j}_1 {\bf j}_2
{\bf s} \rangle$ where we have put ${\bf s}_1 = {\bf s}_2 = {\bf s}$
since $V_2$ conserves spin, parity, and isospin. The number $a$ of such
two--body matrix elements within a major shell is limited. For the
$sd$--shell, we have $a = 63$, while for a single $j$--shell and
identical nucleons, $a = j + 1/2$. For brevity, we denote these matrix
elements by $v_\alpha$ with $\alpha = 1, \ldots, a$. We use the label
$\alpha$ also for the specific two--body operator the two--body matrix
element of which is $v_\alpha$, see~Appendix~1. Within a major shell,
$V_2$ is characterized completely by the $a$ matrix elements
$v_\alpha$, no matter how complicated the actual form of $V_2$.

The Hamiltonian of the shell model is linear in the matrix elements
$v_\alpha$ and has the form
\be
H_{\mu \nu}({\bf J}) = \sum_\alpha v_\alpha C_{\mu \nu}({\bf J},
\alpha) \ .
\label{1}
\ee
The matrices $C_{\mu \nu}({\bf J}, \alpha)$ transport the two--body
interaction into the Hilbert space ${\cal H}({\bf J})$ and depend upon
the quantum numbers ${\bf J}$, upon the particular states $\mu$ and
$\nu$, and upon the particular two--body operator $\alpha$ considered.
The values of the $C_{\mu \nu}({\bf J}, \alpha)$'s are completely
specified by the underlying shell model, i.e., the single--particle
states that occur within the given major shell, the coupling scheme
used to construct the many--body states $| {\bf J} \mu \rangle$, and
by the exclusion principle. The matrices $C_{\mu \nu}({\bf J}, \alpha)$
do not depend upon the choice of the two--body interaction. That choice
is specified by the values of the matrix elements $v_\alpha$.
Eq.~(\ref{1}) yields a decomposition of $H$ into parts which are
determined entirely by the symmetries of the shell model (the matrices
$C_{\mu \nu}({\bf J}, \alpha)$), and parts which carry the information
on the specific details of the two--body interaction (the matrix
elements $v_\alpha$).

We aim at generic statements about spectral properties of $H$ which
apply to (almost) all two--body residual interactions. To this end we
employ the TBRE: The matrix elements 
$v_\alpha$ are assumed to be uncorrelated Gaussian--distributed random
variables with mean values zero and a common second moment $v^2$.
Without loss of generality we put $v^2 = 1$ (we recall that all
single--particle energies are equal so that the scale of the spectrum
is determined by $v^2$). Mean values of observables are worked out
by integrating over the random variables $v_\alpha$, the measure being
given by the product of the differentials of the $v_\alpha$'s and a
Gaussian factor $\exp( - \sum_\alpha v^2(\alpha) / 2)$. Having
calculated the mean value and the square root of the variance of an
observable, we are sure that (within the error given by the latter)
the mean value applies to all members of the ensemble, i.e., to all
two--body interactions, with the exception of a set of measure zero. 
With the $v_\alpha$'s Gaussian random variables, the Hamiltonian
$H_{\mu \nu}({\bf J})$ represents an ensemble of Gaussian--distributed
random matrices, the TBRE.

As mentioned in the Introduction, numerical studies have shown that
the spectral fluctuation properties of the TBRE generically coincide
with those of the GOE. That result is tantamount to saying that $H$
mixes the basis states $| {\bf J} \mu \rangle$ completely, irrespective
of the specific choice of the $v_\alpha$'s, and, thus, reflects a
property of the matrices $C_{\mu \nu}({\bf J}, \alpha)$: Almost every
linear combination of these matrices must be a sufficiently dense
matrix in Hilbert space ${\cal H}({\bf J})$, with matrix elements which
are sufficiently complex, to achieve such mixing. This statement is
rather remarkable because the matrices $C_{\mu \nu}({\bf J}, \alpha)$
are defined entirely in terms of an independent--particle model (which
is integrable). In principle, the $C_{\mu \nu}({\bf J}, \alpha)$'s can
be worked out using group--theoretical methods! Intuitively, that mixing
property of the $C_{\mu \nu}({\bf J}, \alpha)$'s can be understood by
observing that each matrix element of $C_{\mu \nu}({\bf J}, \alpha)$
contains sums of products of Clebsch--Gordan and Racah coefficients and
coefficients of fractional parentage. For more than three particles in
a major shell, the combination of these coefficients is highly complex,
even though each one of them is well--defined and simple. A more
detailed discussion of the properties of the matrices $C_{\mu \nu}({\bf
J}, \alpha)$ was given in Ref.~\cite{PW}. Some further properties are
displayed in Appendix~2. We show, in particular, that the $C_{\mu \nu}
({\bf J}, \alpha)$'s do not commute.

Before exploring some properties of the TBRE in the next Section, we
compare the TBRE with the GOE. To this end, we recall some properties
of the GOE, see Eq.~(\ref{GOE}). The number of independent matrix
elements of the GOE is $N (N+1) / 2$. That is also the number of
uncorrelated random variables which is, thus, very much larger than
the dimension $N$ of the matrices if $N \gg 1$. Except for the symmetry
of the GOE Hamiltonian $H_{\rm GOE}$ about the main diagonal, each
matrix element carries an independent random variable. The situation is
very different for the TBRE. Here, the number $a$ of independent random
variables is typically much smaller than the dimension $d({\bf J})$ of
Hilbert space. The complete mixing of the basis states and the ensuing
validity of Wigner--Dyson statistics for the spectrum cannot be achieved
by such a small number of random variables alone. In an essential way it
is also due to the matrices $C_{\mu \nu}({\bf J}, \alpha)$. These
matrices form a fixed ``scaffolding'' for the TBRE. Although determined
group--theoretically, these matrices are sufficiently dense and complex
to guarantee complete mixing of Hilbert space for almost all choices of
the random variables $v_\alpha$.

The GOE is invariant under orthogonal transformations. Moreover, it is
universal (i.e., the local spectral fluctuation properties do not
depend upon the assumption that the ensemble has a Gaussian
distribution) and ergodic (i.e., in the limit $N \to \infty$ and for
almost all members of the ensemble, the running average of an observable
taken over the spectrum of a single member of the ensemble is equal to
the ensemble average of that observable taken at fixed energy). These
properties make the GOE to a formally attractive, mathematically
accessible and universal model for stochastic spectral fluctuations.
Does the TBRE possess any of these properties? The TBRE is manifestly
not invariant under orthogonal transformations since the matrices
$C({\bf J}, \alpha)$ are fixed by the underlying shell model. It is not
clear whether the TBRE is universal. One might think, for instance, of
modifying the TBRE in such a way that it incorporates known features of
the two--body interaction. One such feature is the pairing force. Doing
so would impose certain fixed relations among the random variables
$v_\alpha$. One could then assume that the remaining unconstrained random
variables have a Gaussian distribution. It is not known whether such a
constrained TBRE would still generically possess spectral fluctuations of
the GOE type. Finally, the ensemble is not ergodic. Indeed, ergodicity
presupposes that the limit of infinite matrix dimension can be taken. But
the matrices $C({\bf J}, \alpha)$ of the TBRE have fixed dimension, and
the question of ergodicity cannot even be formulated meaningfully in this
context. The only exception is the model of a single $j$--shell for which
the limit $j \to \infty$ is meaningful (it shares this property with
French's embedded ensembles). It is not known whether in that limit, the
TBRE for a single $j$--shell is ergodic.

From all these points of view, the GOE is clearly the more attractive
model for stochastic spectral fluctuations. Why then bother with the
TBRE which is mathematically so much more complicated and unrewarding?
The TBRE offers the singular advantage of combining stochastic modelling
with a realistic description of nuclear spectral properties. This leads
to specific insights which go beyond the realm of canonical
random--matrix theory. Some of these are explored in the sequel.

\section{Properties of the TBRE}
\label{prop}

For a full theoretical treatment of the TBRE, we would need a thorough
analytical understanding of the properties of the matrices $C_{\mu
\nu}({\bf J}, \alpha)$. Unfortunately, we are still far from this goal.
We confine ourselves in this Section to a number of results which
illuminate some properties of the TBRE.

\subsection{Information Content of Nuclear Spectra}
\label{info}

Almost by definition, a GOE spectrum carries no information beyond
the symmetry of the Hamiltonian about the main diagonal. Indeed,
the $N (N+1) / 2$ matrix elements of the Hamiltonian are drawn at
random from a Gaussian probability distribution. Conversely, how many
pieces of information are needed to reconstruct the GOE Hamiltonian
from the spectrum? Knowledge of all the $N$ eigenvalues does not
suffice. Knowledge of one eigenfunction adds $N-1$ additional pieces
of information (all expansion coefficients with respect to a fixed
basis except for the constraint due to normalization); knowledge of
a second eigenfunction adds another $N-2$ such pieces (all expansion
coefficients constrained by normalization and orthogonality to the
first eigenfunction). Continuing that way, we see that the knowledge
of all eigenvalues and of all eigenfunctions is needed to reconstruct
the original GOE Hamiltonian. Since a GOE Hamiltonian does not carry
any information, it is useless to investigate a GOE spectrum
spectroscopically.

Can we conclude from this result that whenever a spectrum displays
Wigner--Dyson statistics, it is useless to apply spectroscopy? In
particular, would such a statement hold for nuclei? If not, which is
the information content of a nuclear spectrum? These are the
questions behind the title of the present Subsection.

The answer to the first two questions is obviously a resounding NO.
A Hamiltonian of the TBRE is specified by the choice of the $a$
random variables $v_\alpha$. We recall that $a$ is typically much
smaller than $d({\bf J})$, the dimension of Hilbert space of the
many--body states with quantum numbers ${\bf J}$. Therefore, the
knowledge of $a \ll d({\bf J})$ independent pieces of information
from a spectrum of states with quantum numbers ${\bf J}$ alone will
suffice to determine not only $H({\bf J})$ but also the Hamiltonian
matrices $H({\bf J'})$ of all other quantum numbers ${\bf J'} \neq
{\bf J}$! (This statement carries a small proviso: All the
eigenvalues $s^2_\alpha({\bf J})$ defined in Eq.~(\ref{3}) below
must differ from zero. Otherwise, additional information from
spectra with ${\bf J}' \neq {\bf J}$ is needed).
Given more than $a$ pieces of data, the system of
equations for the $a$ matrix elements $v_\alpha$ is obviously
overdetermined, and nuclear spectroscopy is eminently meaningful even
though the spectra may display Wigner--Dyson statistics. This
conclusion is not confined to low excitation energies but applies,
in principle, also to the domain of slow neutron resonances. The
reason is that the mixing of basis states which leads to
Wigner--Dyson statistics, is mainly due to the scaffolding matrices
$C_{\mu \nu}({\bf J}, \alpha)$ which are fixed by the shell model.

Can we identify the $a$ independent pieces of information which are
best suited to determine the variables $v_\alpha$? More importantly,
do such $a$ independent pieces of information always exist? To answer
these questions, it is useful to rewrite the TBRE in another form. We
use the fact that such questions are most easily addressed in the
framework of linear independence and orthogonality. The matrices 
$C({\bf J}, \alpha)$ span a linear space ${\cal S}$. In ${\cal S}$,
the Hamiltonian~(\ref{1}) can be viewed as a vector with components
$v_\alpha$. The space ${\cal S}$ can be endowed with a scalar product
defined by the normalized trace. For two matrices $C({\bf J}, \alpha)$
and $C({\bf J}, \beta)$, that scalar product has the form
\be
S_{\alpha \beta}({\bf J}) = d^{-1}({\bf J}) \  {\rm Trace} [
C({\bf J}, \alpha) C({\bf J}, \beta) ] \ . 
\label{2}
\ee
The definition~(\ref{2}) has been used in a similar context
before~\cite{FR}. With $\alpha, \beta = 1,2,\ldots,a$, the scalar
products $S_{\alpha \beta}({\bf J})$ of all the $C({\bf J}, \alpha)$'s
form a matrix of dimension $a$. (We recall that in contradistinction,
the $C({\bf J}, \alpha)$'s themselves are matrices in the
$d({\bf J})$--dimensional Hilbert space ${\cal H}({\bf J})$ spanned by
the many--body states $| {\bf J} \mu \rangle$). The $C({\bf J},
\alpha)$'s are real. The cyclic invariance of the trace then implies
that the matrix $S_{\alpha \beta}({\bf J})$ is real and symmetric.
Moreover, the matrix $S_{\alpha \beta}({\bf J})$ is positive
semidefinite. Therefore, $S_{\alpha \beta}({\bf J})$ can be
diagonalized by a real orthogonal matrix ${\cal O}_{\alpha \beta}({\bf
J})$,
\be
({\cal O}({\bf J}) S({\bf J}) ({\cal O}({\bf J}))^T)_{\alpha \beta}
= \delta_{\alpha \beta} \ s^2_\alpha({\bf J}) \ .
\label{3}
\ee
The eigenvalues $s^2_\alpha({\bf J})$ are non--negative and are
arranged by decreasing magnitude, $s^2_1({\bf J}) \geq s^2_2({\bf J})
\geq \ldots s^2_a({\bf J}) \geq 0$. We define the matrices
\be
\overline{B}_{\mu \nu}({\bf J}, \alpha) = \sum_{\beta = 1}^a
{\cal O}_{\alpha \beta}({\bf J}) C_{\mu \nu}({\bf J}, \beta)
\label{4}
\ee
and write Eqs.~(\ref{3}) and ({\ref{2}) as
\be
\frac{1}{d({\bf J})} {\rm Trace} [ \overline{B}({\bf J}, \alpha)
\overline{B}({\bf J}, \beta) ] = \delta_{\alpha \beta} \ s^2_\alpha({\bf
J}) \ . 
\label{3a}
\ee
For all $s^2_\alpha({\bf J})$ with  $s^2_\alpha({\bf J}) = 0$,
Eq.~(\ref{3a}) implies that the trace norm of $\overline{B}({\bf J},
\alpha)$ vanishes which in turn implies that $\overline{B}({\bf J},
\alpha) \equiv 0$. In other words, there exist linear combinations of
the matrices $C({\bf J}, \alpha)$ which vanish identically. We denote
the number of non--vanishing eigenvalues $s^2_\alpha({\bf J})$ by
$a_1$, their positive square roots by $s_\alpha({\bf J})$. The
linear space ${\cal S}$ then has dimension $a_1$. In ${\cal S}$, we
define the $a_1$ orthonormal matrices
\be
B_{\mu \nu}({\bf J}, \alpha) = \frac{1}{s_\alpha({\bf J})}
\sum_{\beta = 1}^a {\cal O}_{\alpha \beta}({\bf J}) C_{\mu \nu}
({\bf J}, \beta) \ {\rm for} \ s^2_\alpha({\bf J}) \neq 0 \ .
\label{4a}
\ee
These are linear combinations of the matrices $C_{\mu \nu}({\bf J},
\beta)$ and obey
\be
\frac{1}{d({\bf J})} \ {\rm Trace} [ B({\bf J}, \alpha) B({\bf J},
\beta) ] = \delta_{\alpha \beta} \ {\rm with} \ \alpha, \beta = 1,2,
\ldots, a_1 \ .
\label{5}
\ee
The matrices $B_{\mu \nu}({\bf J}, \alpha)$ form an orthonormal
basis for the space ${\cal S}$. Using these definitions, we rewrite
the Hamiltonian of the TBRE as
\be
H_{\mu \nu}({\bf J}) = \sum_{\alpha = 1}^{a_1} w_\alpha({\bf J})
s_\alpha({\bf J}) B_{\mu \nu}({\bf J}, \alpha) \ .
\label{6}
\ee
The random variables $w_\alpha(\bf J)$ are linear combinations of
the $v_\alpha$'s,
\be
w_\alpha({\bf J}) = \sum_{\beta = 1}^a {\cal O}_{\alpha \beta}({\bf
J}) v_\alpha \ .
\label{7}
\ee
The $w_\alpha(\bf J)$'s depend upon ${\bf J}$ via the orthogonal
transformation ${\cal O} ({\bf J})$. Like the $v_\alpha$'s, the
$w_\alpha({\bf J})$'s are uncorrelated, have mean values zero and a
common second moment equal to unity. The form~(\ref{6}) of the TBRE
is totally equivalent to the original form~(\ref{1}). The number of
variables $w_\alpha({\bf J})$ appearing explicitly in the
Hamiltonian~(\ref{6}) is $a_1$ and, thus, smaller than $a$ if one or
more of the eigenvalues $s^2_\alpha({\bf J})$ vanish. We shall see
below that that is always the case. According to Eq.~(\ref{6}) the
trace norm of the Hamiltonian $H({\bf J})$ is given by
\be
\frac{1}{d({\bf J})} \ {\rm Trace} [ H^2({\bf J}) ] = \sum_{\alpha
= 1}^{a_1} w^2_\alpha({\bf J}) s^2_\alpha({\bf J}) \ .
\label{8}
\ee
This expression is a measure of the spectral width of the spectrum
of $H({\bf J})$. The spectral width is seen to be a random variable
itself, with mean value $\sum_\alpha s^2_\alpha({\bf J})$. This last
fact offers a physical interpretation of the eigenvalues
$s^2_\alpha({\bf J})$ each of which gives the average spectral width
of the spectrum due to the random variable $w_\alpha({\bf J})$.

Except for accidental degeneracies among the non--vanishing
eigenvalues $s^2_\alpha({\bf J})$, the matrices $B_{\mu \nu}({\bf
J}, \alpha)$ are uniquely defined in terms of the diagonalization
of $S_{\alpha \beta}({\bf J})$. The orthonormality of these
matrices is, of course, unchanged under orthogonal transformations
of the linear space ${\cal S}$, and it is possible to consider
different forms of $H$ arising from such transformations. Among
these, the form~(\ref{6}) is priviledged because the roles of the
random variables $w_\alpha({\bf J})$ and of the eigenvalues
$s^2_\alpha({\bf J})$ are well separated. In all other forms, these
two entities become mixed by the orthogonal transformation.
Nevertheless, there exists another form of $H$ which is physically
interesting. This form is presented in Appendix~3.

We return to the question raised at the beginning of this Subsection:
Can we determine the variables of the Hamiltonian~(\ref{1}) from a
suitable set of data? With $H$ given in the form of Eq.~(\ref{6}),
the relevant variables are now the $w_\alpha({\bf J})$'s. Since the
matrices $B_{\mu \nu}({\bf J}, \alpha)$ have trace norm unity, the
answer depends obviously on the eigenvalues $s^2_\alpha({\bf J})$.
If one or several of these quantities vanish, the Hamiltonian of the
TBRE will, in fact, depend on a smaller number of random variables
than its form~(\ref{1}) suggests. And if one or several of the
$s^2_\alpha({\bf J})$'s are much smaller than the largest, then their
influence upon the spectrum and eigenfunctions will be small, and it
will be difficult or impossible to determine the values of the
associated $w_\alpha$'s from a data set with experimental errors. We
are, thus, led to study the distribution of the eigenvalues
$s^2_\alpha({\bf J})$ or, equivalently, of their positive square
roots $s_\alpha({\bf J})$.

We have calculated the square roots $s_\alpha({\bf J})$ of the
eigenvalues $s^2_\alpha({\bf J})$ numerically for several cases:
For all $J$--values for $m = 6, m = 7$ and $m = 8$ identical nucleons
in the $j = 19/2$ shell, for all $J$--values of the $T = 0$ states in
$^{20}$Ne, and for all $J$--values for the $T = 0$ states in
$^{24}$Mg. We first address the case of the single $j = 19/2$ shell.
Figures~\ref{fig1} and \ref{fig2} show the 
numerical results for the square roots
$s_\alpha(J)$ for $m = 6$ and for $m = 8$ identical nucleons,
respectively. We observe that for all values of $J$, $s_1(J)$ is
distinctly larger than all other roots, and changes very slowly with
$J$, increasing almost monotonically with increasing $J$. The same
statement (near independence of $J$) applies to the corresponding
eigenfunction ${\cal O}_{1 \alpha}(J, m)$. We have calculated the
overlap function $h(m, m'; J, J') = \sum_\alpha {\cal O}_{1 \alpha}
(J, m) {\cal O}_{1 \alpha}(J', m')$ both for $m = m'$ and $J \neq J'$
and for $J = J'$ but $m \neq m'$. Results for the first case were
shown in Fig.~4 of Ref.~\cite{PW1}. These demonstrate that the
overlap function decreases very slowly with increasing distance $|J
- J'|$. Even for $|J - J'| = 20$, the overlap function is still larger
than 0.8. As for the second case, we have calculated the overlap
function for $J = J' = 0,1,2,\ldots,10$ and $(m, m') = (6, 8)$, for
$J = 0,1,2,\ldots,10$ with $J' = J+1/2$ and $(m, m') = (6, 7)$, and
for $J' = 0,1,2,\ldots, 10$ with $J = J' + 1/2$ and $(m, m') = (7,
8)$. The overlap function is larger than $0.995$ in all cases except
for pairs involving $m = 6, J = 1$. This last fact is due to a special
feature of the matrix $S_{\alpha \beta}$ for $m = 6, J = 1$: Two
eigenvalues of $S_{\alpha \beta}$ are zero. We conclude that the
matrices $B_{\mu \nu}(J, m, \alpha = 1)$ consist of (nearly) the same
linear combinations of the matrices $C_{\mu \nu}(J, m, \alpha)$ for
many values of $J$ and $m$.

\begin{figure}[h]
\includegraphics[width=0.45\textwidth]{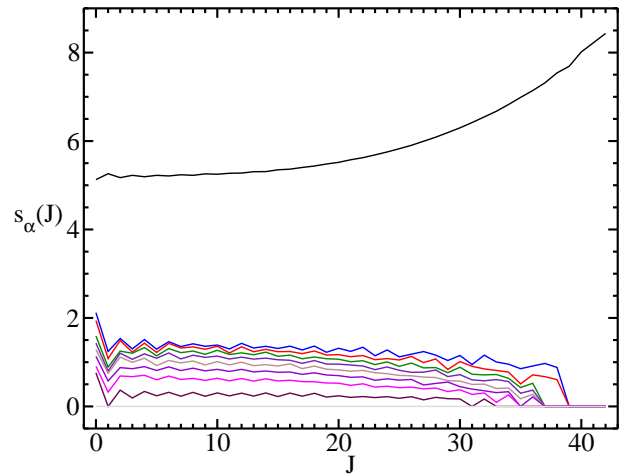}
\caption{\label{fig1}(Color online) The square roots of the eigenvalues
of the matrix $S_{\alpha \beta}(J)$ for a single $j$--shell with
$j = 19/2$ and $m = 6$ identical nucleons versus total spin $J$.}
\end{figure}

\begin{figure}[h]
\includegraphics[width=0.45\textwidth]{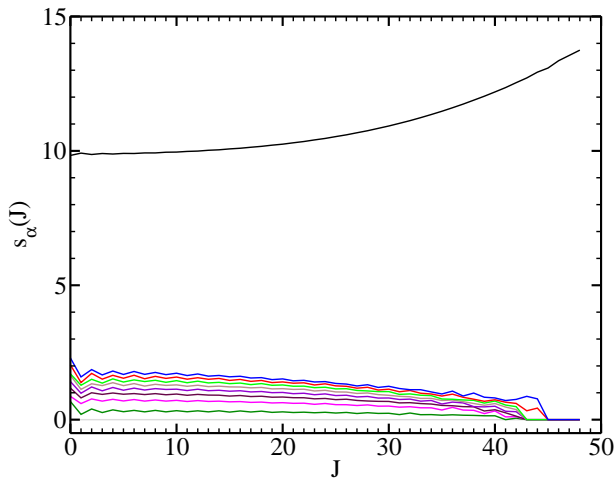}
\caption{\label{fig2}(Color online)Same as Fig.~\ref{fig1} but for 
$m = 8$ nucleons.}
\end{figure}

Some of these features can be understood analytically via arguments
presented in Appendix~4. We use the numerically established near
independence of the largest eigenvalue $s_1(J)$ and of the associated
eigenfunction ${\cal O}_{1 \alpha}(J)$ on $J$ to study the average
$X_{\alpha \beta}$ of $S_{\alpha \beta}$ (averaged over all
$J$--values). We do so in the limit $(2j+1) \gg 1$. In this limit,
$X_{\alpha \beta}$ factorizes, $X_{\alpha \beta} \approx f(\alpha)
f(\beta)$. Hence all eigenvalues but one vanish, and $s^2_1$ is given
by $\sum_\alpha f^2(\alpha)$. For $m = 6$ ($m = 8$) this yields $s_1
\approx 6.3$ ($s_1 \approx 11.4$, respectively). We also determine
the eigenfunction ${\cal O}_{1 \alpha}$ which (except for
normalization) has components $(1,5,9,\ldots)$. The resulting matrix
$B_1$ is
\be
\label{monopole}
B_1 \propto \sum_s (2s+1) C({\bf J}, 1+s/2) \ .
\ee
The sum runs over $s = 0,2,4,\ldots,2j-1$ for the values of the
two--particle spin. The matrix $B_1$ is the matrix of the monopole
operator ${\cal B}_1$. Inspection reveals that this operator is
approximately proportional to the unit operator. Thus, all matrices
that are orthogonal (under the trace) to $B_1$ must have approximately
zero trace. Moreover, the leading term in the transformed
Hamiltonian~(\ref{6}) is approximately a scalar. It defines the
approximate value of the centroid of the eigenvalues and has
presumably only a small influence on the spectral statistics. The
value of the centroid is determined by the values of $s_1({\bf J})$
and of $w(1)$. The spectral statistics are largely determined by the
remaining terms in $H$.

Our analytical results are mirrored in the numerical findings. Except
for the largest $J$--values, the matrices $B(J,1)$ are close to the
unit matrix in $d(J)$ dimensions. Comparison with Figs.~\ref{fig1} 
and \ref{fig2} 
shows that the analytical results for the eigenvalues are
semiquantitatively correct. For the eigenfunction ${\cal O}_{1
\alpha}$ the numerical diagonalization for $m = 6$ and $J = 0$ yields
an overlap of 0.97 with the analytical result.

We turn to the data for the $sd$--shell. Our calculations are partly
based on the shell--model code Oxbash \cite{Ox}. The square roots
$s_\alpha(J)$ for $^{20}$Ne are shown in Fig.~\ref{fig3}, those for
$^{24}$Mg are shown in Fig.~\ref{fig4} The tendencies are
qualitatively similar to those in the single $j$--shell. In $^{20}$Ne,
$s_1(J)$ is not larger than all other roots by as big a factor as in
the single $j$--shell, or as in $^{24}$Mg. This suggests that the
clear separation of $s_1(J)$ from the rest of the roots requires a
minimum number of nucleons. Indeed, in Appendix~4, the factorization
of $X_{\alpha \beta}$ occurs only for $m \geq 4$, and $s_1$ grows with
$m$ roughly like $m^2$. Thus, our results suggest a generic behavior
for many nucleons in the shell. Again, we find that the leading matrix
$B(J, 1)$ is (except for a normalization factor) approximately equal
to the unit matrix.

\begin{figure}[h]
\includegraphics[width=0.45\textwidth]{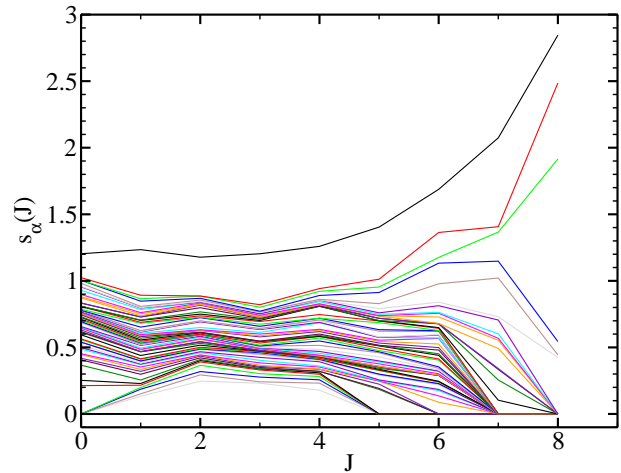}
\caption{\label{fig3}(Color online)  
The square roots of the eigenvalues
of the matrix $S_{\alpha \beta}(J)$ for the $sd$--shell nucleus
$^{20}$Ne with isospin $T = 0$ versus total spin $J$.}
\end{figure}

\begin{figure}[h]
\includegraphics[width=0.45\textwidth]{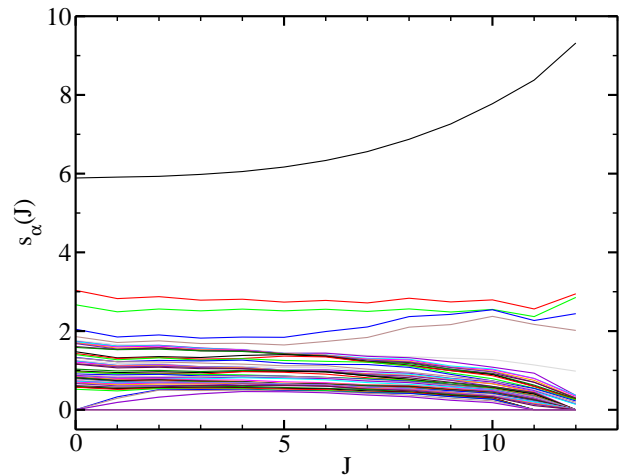}
\caption{\label{fig4}(Color online) Same as Fig.~\ref{fig3} but for 
$^{24}$Mg.}
\end{figure}

Another remarkable feature in all the figures is the disappearance of
one or several roots for some values of $J$. In fact, for each spin, at
least one eigenvalue vanishes identically. The corresponding scalar
two--body operator is given by ${\hat J}^2 - J(J+1)$ where ${\hat J}$
denotes the operator of total spin. The matrix representation of this
operators is a linear combination of the matrices $C_{\mu \nu}({\bf
J})$. In the case of the $sd$--shell nuclei, an analogous statement
applies to the isospin operator ${\hat T}^2 - T(T+1)$. More eigenvalues
than one vanish in cases where the number of independent matrix elements
of $H({\bf J})$ is smaller than the number of two--body matrix elements.

How do these results relate to the actual determination of the
effective shell model interaction from experimental data? We
recall that usually one proceeds in two steps. In a first step, the
effective interaction is determined through a $G$--matrix calculation
that is based on nucleon--nucleon potentials which fit
nucleon--nucleon scattering data with very high accuracy. The
resulting $G$--matrix usually does not reproduce experimental nuclear
spectra well enough, and corrections have to be made in a second
step. There are basically two ways to make these corrections. A first
approach insists on minimal corrections and only corrects the monopole
term by a fit to experimental data. This is very reasonable since the
monopole term has the largest trace norm, and is essentially the only
matrix with a finite trace (or centroid). This last property of the
monopole has been known since the pioneering work of Pasquini and
Zuker \cite{Pas78}. We note that the resulting shell--model
Hamiltonian has an impressive predictive power, see, e.g., the
shell--model studies of $fp$--shell nuclei by the Strasbourg
group~\cite{Cau99}. In the second and alternative approach, one
attempts to fit all two--body matrix elements to experimental data.
This is done through an iterative procedure that starts from the
$G$--matrix and makes small changes in the matrix elements until a
best fit is obtained. As the number of data points (i.e., energy
levels) usually exceeds the number of two--body matrix elements, the
problem is over--determined, and only a smaller number of linear
combinations of data points can be used in the fit due to
rank--deficiency problems of the matrix equations
involved~\cite{Hon02}. As an example we mention the widely used
Brown--Wildenthal interaction~\cite{Bro88} for $sd$--shell nuclei
where only 47 linear combinations of the 66 parameters (63 two--body
matrix elements and three single--particle energies) are determined by
fit while the remaing ones are taken from the $G$--matrix. In larger
model spaces the ratio of well--determined linear combinations usually
decreases. For the $fp$--shell, only 70 linear combinations out of 199
free parameters are used in the fit to data~\cite{Hon04}. As shown by
our results in this Subsection, the answer to the question ``Which are
the most important linear combinations of two--body matrix elements''
is less determined by the data used in the fit but rather by the
inherent structure of the shell model itself. Our work identifies
which linear combinations these are: The $w_\alpha({\bf J})$'s which
correspond (for fixed ${\bf J}$) to the largest
eigenvalues. Unfortunately, the property of an eigenvalue to be large
or small changes little (if at all) with ${\bf J}$, and the same is
true of the orthogonal transformation defining the random variables
$w_\alpha({\bf J})$. This fact shows that even an enlarged data set
might not necessarily lead to more accurate values for the
$w_\alpha({\bf J})$'s.

In conclusion, the information content of a nuclear spectrum as given by
the TBRE is very different from that of a GOE. The actual possibility
to determine the unknown parameters $w_\alpha({\bf J})$ depends on the
magnitudes of the eigenvalues $s^2_\alpha({\bf J})$. We stress that
these eigenvalues are uniquely determined by the underlying shell model
and do not depend upon the residual interaction. Thus, they may be
calculated prior to any attempt to fit the residual interaction to
actual data. We believe that such a procedure might be useful for
practical shell--model work.

\subsection{Preponderance of Ground States with Spin Zero}
\label{prep}

In 1998, Johnson, Bertsch, and Dean~\cite{JBD} found that the TBRE is
likely to yield spin--zero ground states in even--even nuclei with a
probability which is considerably larger than the fraction of states
with spin zero in the total shell--model space. This came as a
surprise because the TBRE does not have a built--in pairing force or
quadrupole force. Subsequent work showed that similar regularities
exist in bosonic~\cite{BiF} and electronic~\cite{JS} many--body
systems with two--body interactions. The phenomenon of spin--zero
preponderance seems a robust and rather generic feature which has
received much attention since, see the reviews~\cite{ZV,ZAY} and
references therein. Here we discuss it in the framework of our
representation~(\ref{6}) of the Hamiltonian of the TBRE. We extend our
earlier work in Ref.~\cite{PW1}.  As before in this paper, we focus
attention on the case of a single $j$--shell and on the $T = 0$ states
of nuclei in the $sd$--shell and, therefore, use the label $J$ rather
than ${\bf J}$.

Our approach is based upon the following consideration. For each value
of $J$ and each realization of the random variables $v_\alpha$, we
define the spectral radius $R(J)$ as the distance of the farthest
eigenvalue from the center of the spectrum. (This can be the smallest
or the largest eigenvalue. In view of the randomness of the signs of
the $w_\alpha(J)$'s, that ambiguity is irrelevant). The $R(J)$'s are
random variables. For every value of $J$ we determine the probability
that the spectral radius $R(J)$ is maximal. To this end we relate
$R(J)$ to the spectral width $\sigma(J)$. The square of the spectral
width $\sigma(J)$ is given by the normalized trace norm of the
Hamiltonian, see Eq.~(\ref{8}),
\be
\sigma^2(J) = \sum_{\alpha = 1}^{a_1} w^2_\alpha(J) s^2_\alpha(J) \ . 
\label{10}
\ee
We postulate a linear relationship between the spectral radius and the
spectral width,
\be
R(J) = r(J) \sigma(J) \ .
\label{12}
\ee
We show that the scaling factors $r(J)$ are nearly constant (i.e.,
independent of the random variables $w_\alpha(J)$) but do depend upon
$J$. Then, Eq.~(\ref{12}) relates the random variables $R(J)$ with the
random variables $\sigma(J)$. The dependence of the scaling factors
$r(J)$ on $J$ and the distribution of and correlations among the
spectral widths $\sigma(J)$ together lead to an understanding of the
preponderance of spin--zero ground states.

We first address the scaling factors $r(J)$. The inset of
Fig.~\ref{fig5} shows the dependence of $R(0)$ on $\sigma(0)$ for 6
identical nucleons in the $j = 19/2$ shell, together with a linear
fit, for 900 realizations of the TBRE. We see that Eq.~(\ref{12})
holds with an approximately constant value of $r(0)$. The reduced
$\chi^2$ per degree of freedom is about 0.9. This value increases
toward 1.5 for spins around $J \approx 20$. Figure~\ref{fig5} shows
the linear--fit values of $r(J)$ versus $J$. These have an overall
tendency to decrease with increasing $J$. This fact reflects the
overall decrease of the number $d(J)$ of states of spin $J$ with
increasing $J$. Indeed, the smaller the number of states in the
spectrum the closer do we expect $R(J)$ and $\sigma(J)$ to be. This
expectation can be quantified. We recall that a shell--model spectrum
with spin $J$ calculated from the TBRE has approximately Gaussian
shape~\cite{MF}. For a Gaussian spectrum with spectral width $\sigma$
normalized to the total number $d$ of states, the average level
density $\rho(E)$ has the form
\be
\rho(E) = \frac{d}{\sqrt{2 \pi} \sigma} \exp( - E^2 / (2 \sigma^2) ) \ .
\label{13}
\ee
The most likely position of the lowest (the highest) state in the
spectrum is found by integrating $\rho(E)$ from $- \infty$ ($+ \infty$,
respectively) to the point where the area under the integral equals
$1/2$. This point is equal to $R$. Thus,
\be
\frac{1}{2} = \frac{d}{\sqrt{2 \pi}} \int_{- \infty}^{- R/\sigma} {\rm
d} x \exp( - x^2 / 2 ) \ . 
\label{14}
\ee
The solutions $R / \sigma = r(d)$ of this equation depend on $d$. In
Figure~\ref{fig5}, we have displayed the resulting values for $r(d(J))$,
again for the $j = 19/2$ shell with $m = 6$ identical nucleons. The
odd--even staggering reflects corresponding changes in the dimensions
$d(J)$. The error in our theoretical determination is expeced to be of
the order of the mean level spacing, i.e., of order $d^{-1}$ and, thus,
small for $d \gg 1$. The actual discrepancy between the data and our
analysis is much larger. We ascribe this to the fact that the assumed
Gaussian shape of the spectrum holds only approximately.

\begin{figure}[h]
\includegraphics[width=0.45\textwidth]{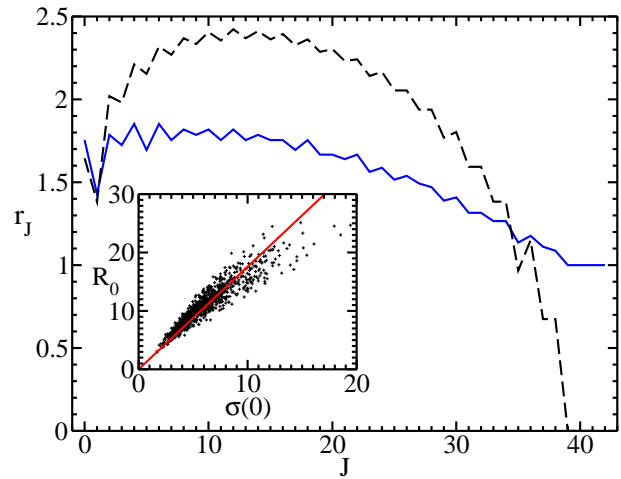}
\caption{\label{fig5}(Color online) Dependence of the scaling factor
$r(J)$ in Eq.~(\ref{12}) on spin $J$ for $m = 6$ identical nucleons in
the $j = 19/2$ shell. Solid line: Data based on the linear fit. Dashed
line: Prediction based on Eq.~(\ref{14}). Inset: Dependence of $R(0)$
on $\sigma(0)$ for 900 random realizations of the TBRE. Solid line: The
linear fit which determines $r(0)$.}
\end{figure}

We turn to the distributions of and the correlations among the spectral
widths $\sigma(J)$. From Eq.~(\ref{10}), the ensemble average (henceforth
denoted by an overbar) of $\sigma^2(J)$ is
\be
\overline{\sigma^2(J)} = \sum_{\alpha = 1}^{a_1} s^2_\alpha(J) \ ,
\label{15}
\ee
and the variance is
\be
\overline{\sigma^4(J)} - \bigg( \overline{\sigma^2(J)} \bigg)^2 = 2
\sum_{\alpha = 1}^{a_1} s^4_\alpha(J) \ .
\label{16}
\ee
The normalized root of the variance equals $\sqrt{2/a_1}$ when all
eigenvalues are equal and equals $\sqrt{2}$ when all eigenvalues but
one vanish. Therefore, we conclude from Figures~\ref{fig1} to \ref{fig4}
that the fluctuations of $\sigma^2(J)$ are biggest for the largest
$J$--values. This is also borne out when we calculate the complete
probability distribution $P_J(\sigma) = 2 \sigma \overline{\delta(
\sigma^2 - \sigma^2(J))}$ for $\sigma(J)$. We write the
$\delta$--function as a Fourier integral and perform the integrations
over all the Gaussian variables $w_\alpha(J)$. We find
\be
P_J(\sigma) = \frac{\sigma}{\pi} \int_{- \infty}^{+ \infty} {\rm d} t
\ e^{i t \sigma^2} \ \prod_{\alpha = 1}^{a_1} \frac{e^{- \frac{i}{2}
{\rm arctan} (2 t s^2_\alpha(J))}}{(1 + 4 t^2 s^4_\alpha(J))^{1/4}} \ .
\label{17}
\ee
Plots of $P_J(\sigma)$ for various values of $J$ were shown in
Figure~3 of Ref.~\cite{PW1}. The flattest curve is the one for the
largest value of $J$ as expected.

Central to an understanding of the preponderance of spin--zero ground
states in nuclei are the correlations between the spectral widths
$\sigma(J)$ pertaining to different values of $J$. We have seen that
both the eigenfunction pertaining to the largest eigenvalue and all
eigenvalues of the matrix $S_{\alpha \beta}$ change little with $J$.
In the extreme case where for a pair $(J, J')$ of spins we would have
$s^2_\alpha(J) = s^2_\alpha(J')$ and $w_\alpha(J) = w_\alpha(J')$ for
all values of $\alpha$, the spectral widths would obviously be totally
correlated: Their joint probability distribution is proportional to a
delta function $\delta (\sigma(J) - \sigma(J'))$. In this case, the
value of the level with the lowest spin depends not on $\sigma(J)$
but only on $r(J)$. Actually the eigenvalues are not exactly equal
and, more importantly, the eigenfunctions for the smaller eigenvalues
differ, and so do the corresponding $w_\alpha(J)$'s. Still, we must
expect significant correlations.

The correlations cannot be worked out directly from Eq.~(\ref{10}) as
this would require knowledge of the correlations among the
$w_\alpha(J)$'s for different values of $J$. Rather, we use that
$\sigma^2$ is defined in terms of the trace norm of the Hamiltonian,
use Eq.~(\ref{1}) for the latter and the definition~(\ref{2}) for
$S_{\alpha \beta}(J)$ and have
\be
\sigma^2(J) = \sum_{\alpha \beta} v_\alpha v_\beta S_{\alpha \beta}(J)
\ .
\label{18}
\ee
For the covariance of two $\sigma^2$'s this yields
\be
\overline{\sigma^2(J) \sigma^2(J')} - \overline{\sigma^2(J)} \
\overline{\sigma^2(J')} = 2 \ {\rm Trace} [S(J) S(J')] \ .
\label{19}
\ee
In general, the right--hand side of this equation differs from zero. It
is not possible straightforwardly to calculate the joint probability
distribution of $\sigma(J)$ and $\sigma(J')$. However, the probability
\be
p(J,J') = \overline{\Theta(\sigma^2(J) - \sigma^2(J'))}
\label{20}
\ee
that spin $J$ has larger spectral width than spin $J'$ can be
calculated in terms of the eigenvalues $q_\alpha$ of the matrix
$(S_{\alpha \beta}(J) - S_{\alpha \beta}(J'))$. The calculation is
quite similar to the one which leads to Eq.~(\ref{17}) and yields
\be
p(J, J') = \frac{1}{2} + \frac{1}{\pi} \int_{0}^\infty \frac{{\rm d} t}
{t} \frac{\sin (\frac{1}{2} \sum_\beta \arctan 2 t q_\beta )}
{\prod_\alpha (1 + 4 t^2 q^2_\alpha)^{1/4}} \ . 
\label{21}
\ee
This formula is in good agreement with our numerical results.

We emphasize that our approximations for $r(J)$ and $\sigma(J)$ do not
yield a faithful representation of the distributions and correlations
of the spectral radii $R(J)$. However, these approximations turn out
to be sufficiently accurate to predict the probability that $R(J) =
r(J) \sigma(J)$ has maximum value. These predictions are compared with
the results of numerical diagonalization in Figs.~\ref{fig6} to
\ref{fig9}. In all four Figures, we plot the probability that the
ground state has spin $J$ versus $J$. The data points are calculated
from 900 realizations of the TBRE. The solid lines show the
probability that the spectral width $\sigma(J)$ has the largest
value. The dashed lines show the probability that the product $r(J)
\sigma(J)$ has the largest value.

\begin{figure}[h]
\includegraphics[width=0.45\textwidth]{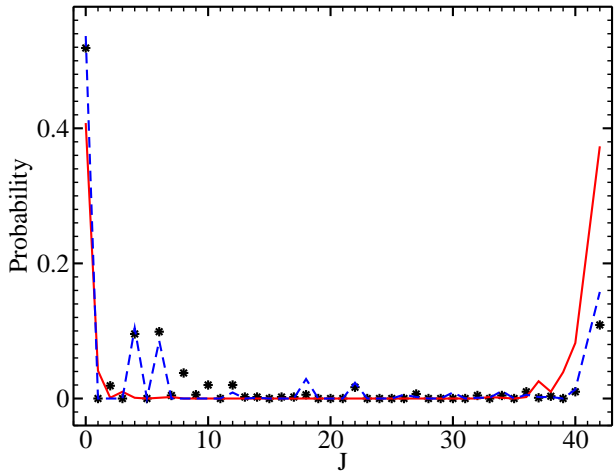}
\caption{\label{fig6}(Color online) The probability that the ground
state has spin $J$ versus $J$ for $m = 6$ identical nucleons in the $j
= 19/2$ shell. Dots: Data points from 900 realizations. Solid line:
Probability that the spectral width $\sigma(J)$ has the largest
value. Dashed line: Probability that the product $r(J) \sigma(J)$ has
the largest value. }
\end{figure}

\begin{figure}[h]
\includegraphics[width=0.45\textwidth]{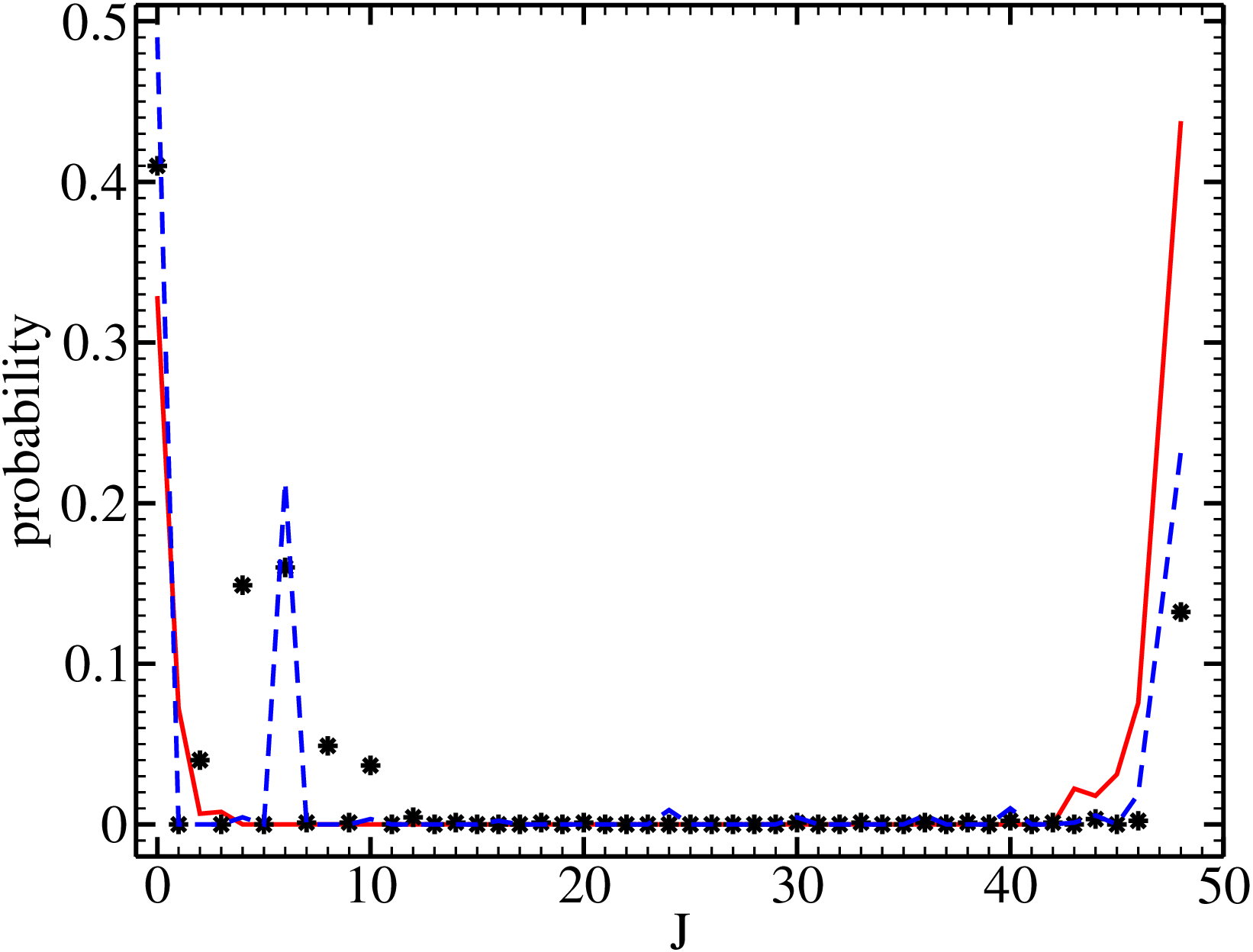}
\caption{\label{fig7}(Color online) Same as Fig.~\ref{fig6} but for $m
= 8$ identical nucleons}
\end{figure}

\begin{figure}[h]
\includegraphics[width=0.45\textwidth]{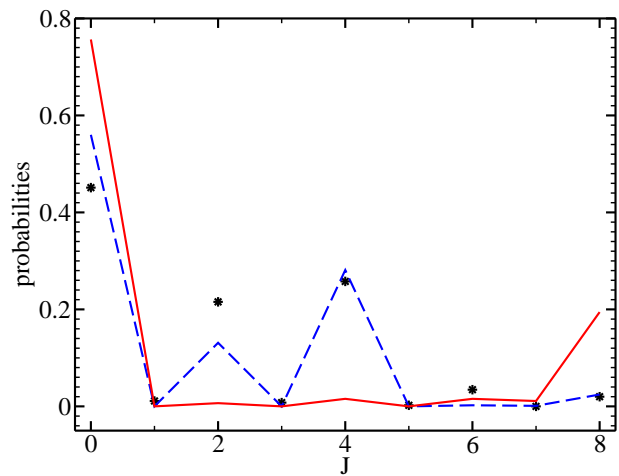}
\caption{\label{fig8}(Color online) Same as Fig.~\ref{fig6} but for
the $T = 0$ states in $^{20}$Ne.}
\end{figure}

\begin{figure}[h]
\includegraphics[width=0.45\textwidth]{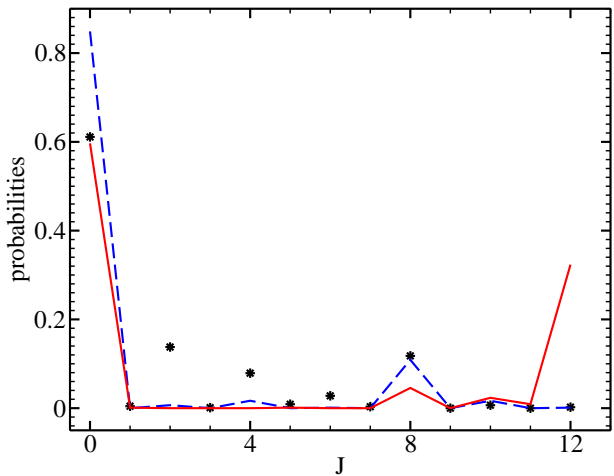}
\caption{\label{fig9}(Color online) Same as Fig.~\ref{fig6} but for
the $T = 0$ states in $^{24}$Mg.}
\end{figure}

We observe that the states with spin zero and with maximum spin have
the largest spectral widths. The factor $r(J)$ suppresses the high
$J$--values. It causes a staggering in the probabilities which is not
present for the spectral widths. In all Figures, the agreement of the
dashed lines with the data points is satisfactory. We note that the
predictions are somewhat less accurate for $^{24}$Mg. A more detailed
analysis of this case shows that the determination of the scale factor
$r(J)$ by fit is somewhat less accurate as the typical $\chi^2$ per
datum is around 2.5.

It is of some interest to see whether our arguments apply also to
nuclei with an odd number of nucleons and half--integer total spin. In
complete analogy to Figs.~\ref{fig6} to \ref{fig9}, we have calculated
the probabilities for the half--integer spin states for $m = 7$
identical nucleons in the single $j = 19/2$ shell, to form the ground
state. The results are shown in Fig.~\ref{fig10}, with symbols that
carry the same meaning as before. Again, the agreement is very
convincing. This strengthens our belief that our approach is generic.

\begin{figure}[h]
\includegraphics[width=0.45\textwidth]{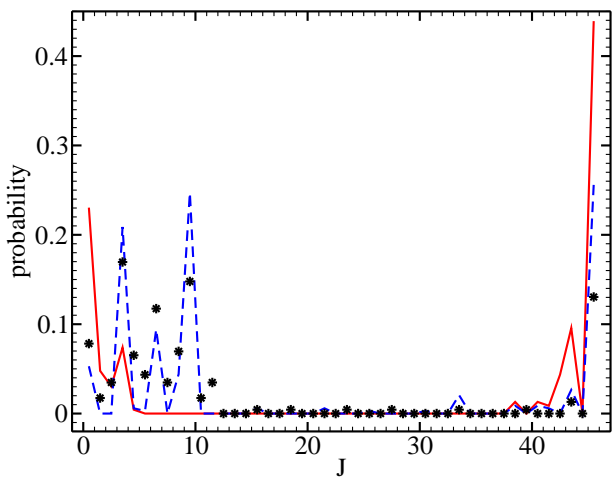}
\caption{\label{fig10}(Color online) Same as Fig.~\ref{fig6} but for $m
= 7$ identical nucleons}
\end{figure}

In summary, we have shown that our semi--analytical approach to the
TBRE offers a satisfactory explanation for the preponderance of ground
states with zero spin observed for this ensemble. We have established
an approximate proportionality between two sets of random variables,
the spectral radii $R(J)$ and the spectral widths $\sigma(J)$, $R(J)
\approx r(J) \sigma(J)$, with nearly constant scaling factors $r(J)$.
We have presented semi--analytical results for $r(J)$, and closed
expressions for the distribution functions for the spectral widths as
well as for some of their correlations and joint distribution functions.
The scaling factors $r(J)$ depend essentially on the dimensions $d(J)$
of the associated Hilbert spaces, show odd--even staggering, and an
overall decrease with increasing $J$. The spectral widths reflect
properties of the underlying shell model and fluctuate most strongly
for large values of $J$. Their average values are largest for $J = 0$
and for maximum $J$. In the products $R(J)$, the large $J$--values are
suppressed by the scaling factors $r(J)$. The good agreement of our
theoretical predictions with the data in all cases suggests that we
have identified the generic mechanism which is responsible for the
preponderance of spin--zero ground states. That mechanism should
likewise operate in other major shells and perhaps in other many--body
systems governed by two--body interactions.

\subsection{Correlations between Spectra with Different \\
Quantum Numbers and/or in Different Nuclei}
\label{corr}

For different quantum numbers ${\bf J} \neq {\bf J}'$ and/or different
nucleon numbers $m \neq m'$, the Hamiltonian matrices of the TBRE
defined in Eq.~(\ref{1}) depend upon the same set of random variables
and are, therefore, correlated. This fact is rather obvious from the
point of view of nuclear physics: Changing the residual interaction
will simultaneously change all the spectra in all the nuclei within
the major shell under consideration. However, from the point of view
of random--matrix theory, the existence of correlations between spectra
each of which displays Wigner--Dyson type level fluctuations, is
excluded by assumption. To the best of our knowledge, a theoretical
framework for the treatment of such correlations does not exist. The
parametric level correlations which have been discussed earlier, in the
context of both condensed--matter theory~\cite{AS} and random--matrix
theory~\cite{W}, do not cover the present case. Indeed, there one
considers a Hamiltonian which depends upon an external parameter like
the strength of an external magnetic field. The dimension of the
Hamiltonian matrix remains unchanged as the parameter is varied. Here,
in contradistinction, correlations exist between Hamiltonian matrices
which differ in matrix dimension. Moreover, parametric level
correlations generically tend to zero as the difference between the old
and the new values of the parameter increases. In the present case, it
is not clear whether the correlation between two spectra differing, say,
by two units of total spin is larger or smaller than that between two
spectra differing by ten units. Also, it is not clear whether such
correlations are big enough to be significant experimentally. These
facts prompt us to explore the magnitude of spectral correlations both
versus spin and versus mass. We focus attention on the case of
$sd$--shell nuclei.

Using the standard definition for the level density
\be
\rho(E,J) = \sum_\mu \delta(E - E_\mu(J))
\label{9}
\ee 
in terms of the eigenvalues $E_\mu(J)$ pertaining to spin $J$, we have
calculated numerically the density--density correlations for $^{24}$Mg
for $T = 0$ states with $J = 0$ and $J = 1$. Our ensemble average uses
400 realizations of the TBRE. The results are shown in
Fig.~\ref{fig11}. The left panel shows as a density plot the mean
value of the product of the level densities, $\overline{\rho(E, 0)
\rho(W, 1)}$, versus the energies $E$ and $W$ in units of $v^{-2}$.
The midddle panel shows analogously the product of the mean densities
$\overline{\rho(E, 0)}$ $\times$ $\overline{\rho(W, 1)}$.  This is
essentially the product of two Gaussian distributions. The right panel
shows the actual correlator, i.e., the difference $\overline{\rho(E,
0) \rho(W, 1)} - \overline{\rho(E, 0)} \times \overline{\rho(W,
1)}$. The maximum value of the correlator is about 10\% of the product
of the mean level densities. This demonstrates the existence of weak
correlations. The correlator has a minimum in the center of the two
spectra. We cannot exclude the possibility that with increasing matrix
dimension, this minimum widens, so that significant correlations would
exist only in the tails of the spectra. We have found correlations
between spectra pertaining to different $J$--values also in the case
of a single $j$-shell, where they are significantly stronger and are
of the order of 50\%. Correlations are also expected to exist between
states in different nuclei when these are governed by the same TBRE. A
case in point are the $T = 0$ spin zero states in $^{20}$Ne and
$^{24}$Mg. Their correlations are shown in Figure~\ref{fig12} where we
use the same display as in Figure~\ref{fig11}. Again, we find weak
correlations of the order of several percent.

\begin{widetext}

\begin{figure*}[h]
\includegraphics[width=0.3\textwidth]{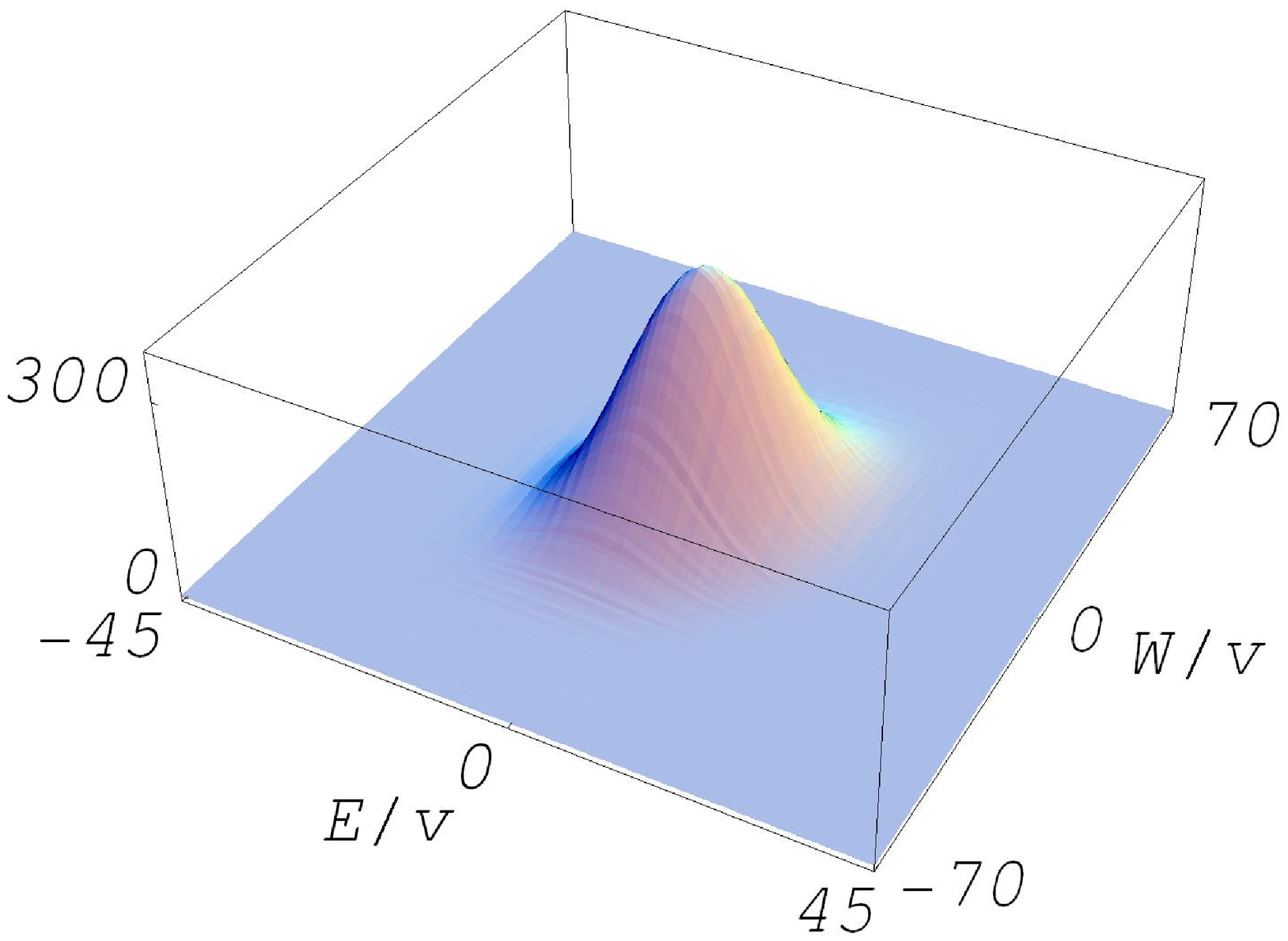}
\includegraphics[width=0.3\textwidth]{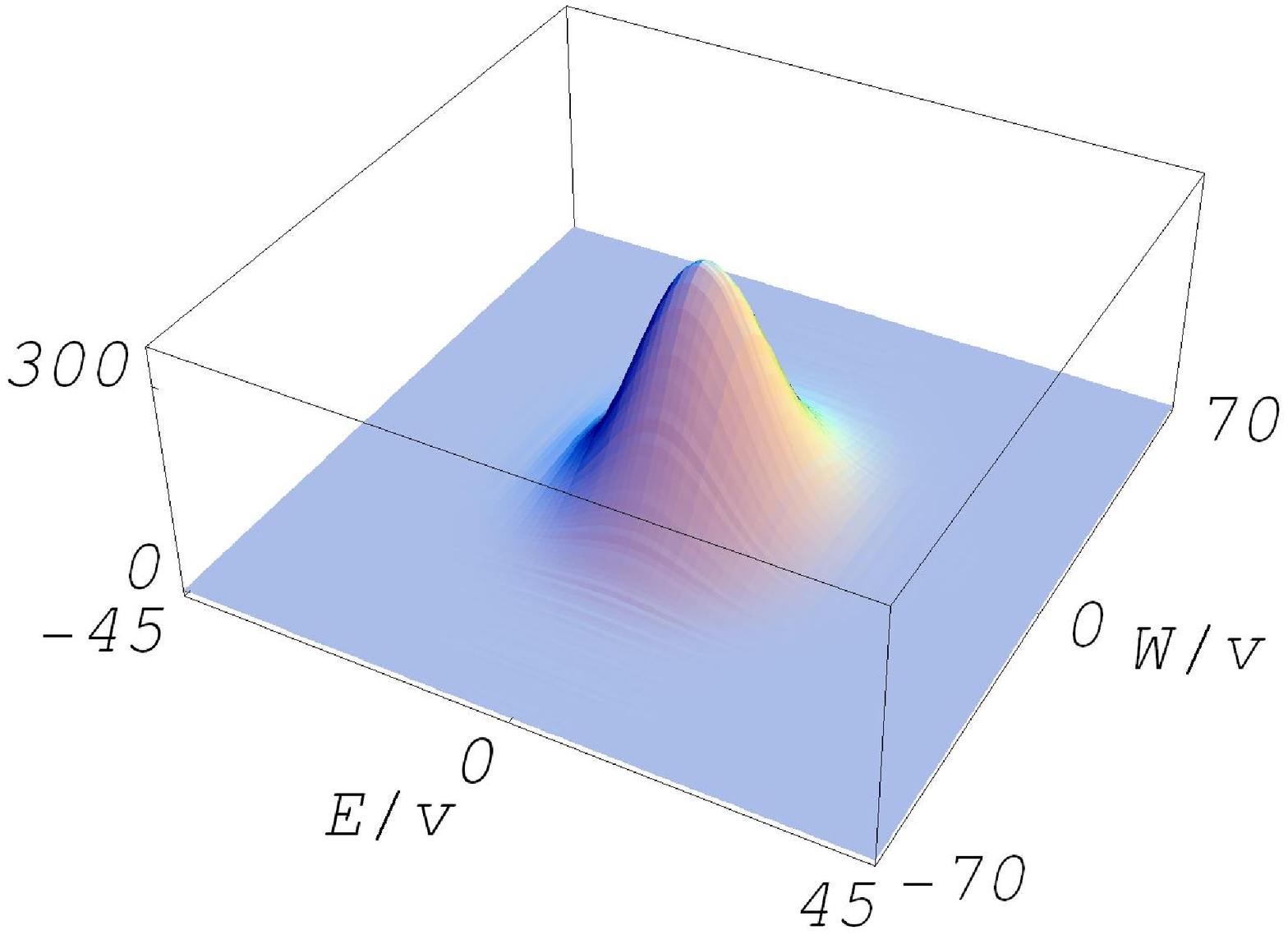}
\includegraphics[width=0.3\textwidth]{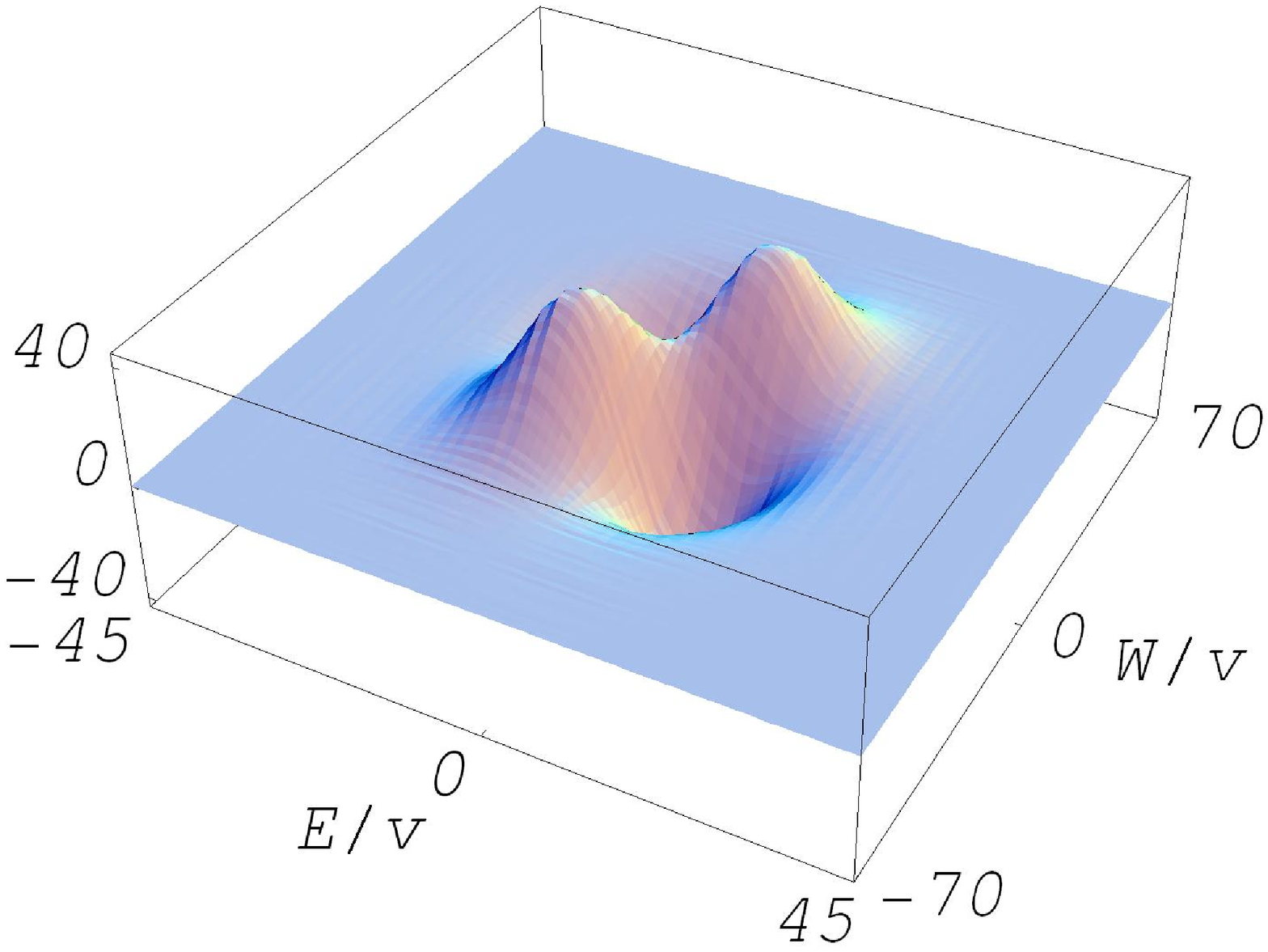}
\caption{\label{fig11} Correlations between the $J=0, T=0$ states and 
the $J=1,T=0$ states of $^{24}$Mg (from 400 realizations of the ensemble).
Left panel: Average of the product of the two level densities. Middle
panel: Product of the averages of the two level densities. Right panel:
The correlator.}
\end{figure*}

\begin{figure*}[h]
\includegraphics[width=0.3\textwidth]{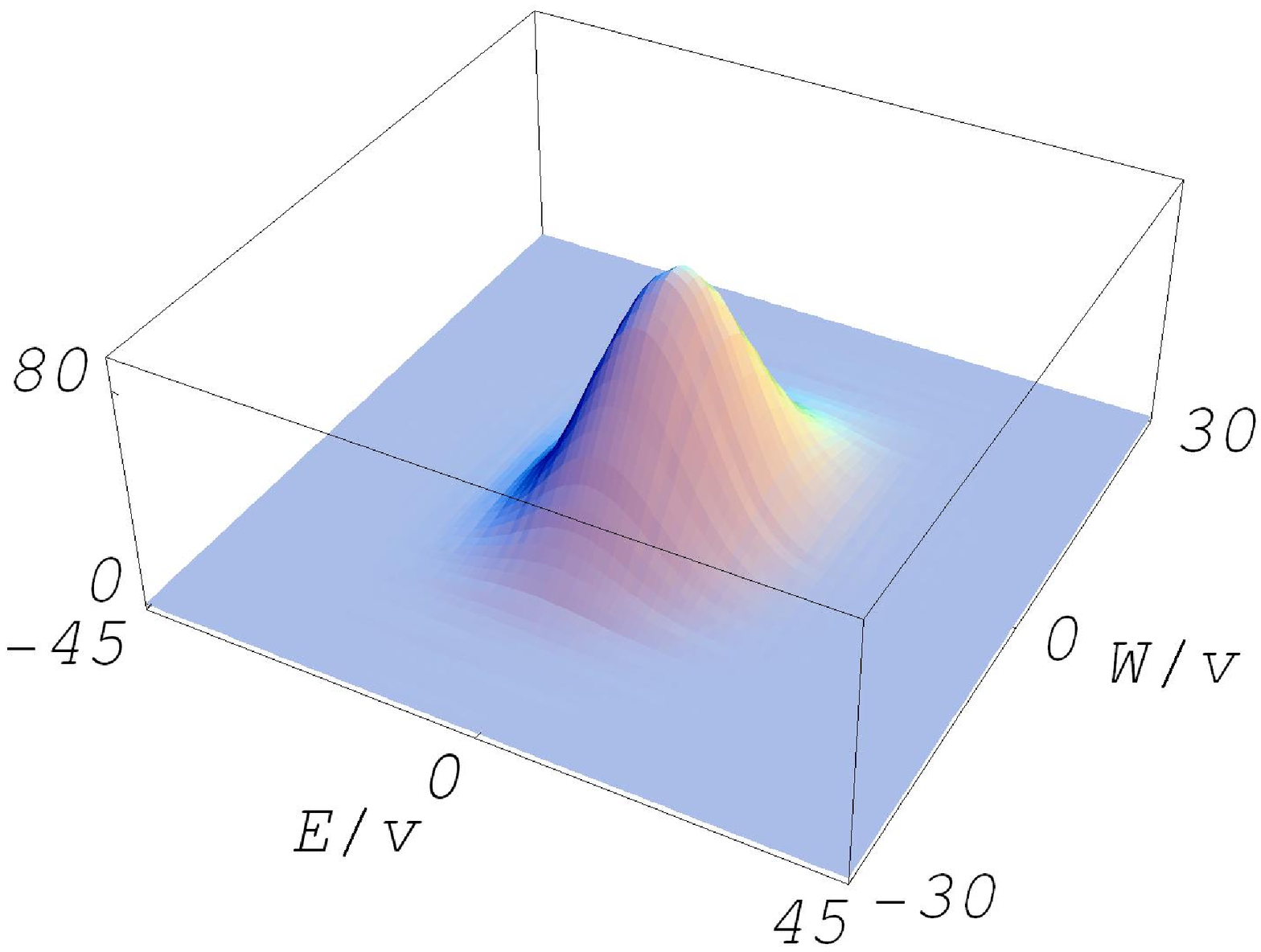}
\includegraphics[width=0.3\textwidth]{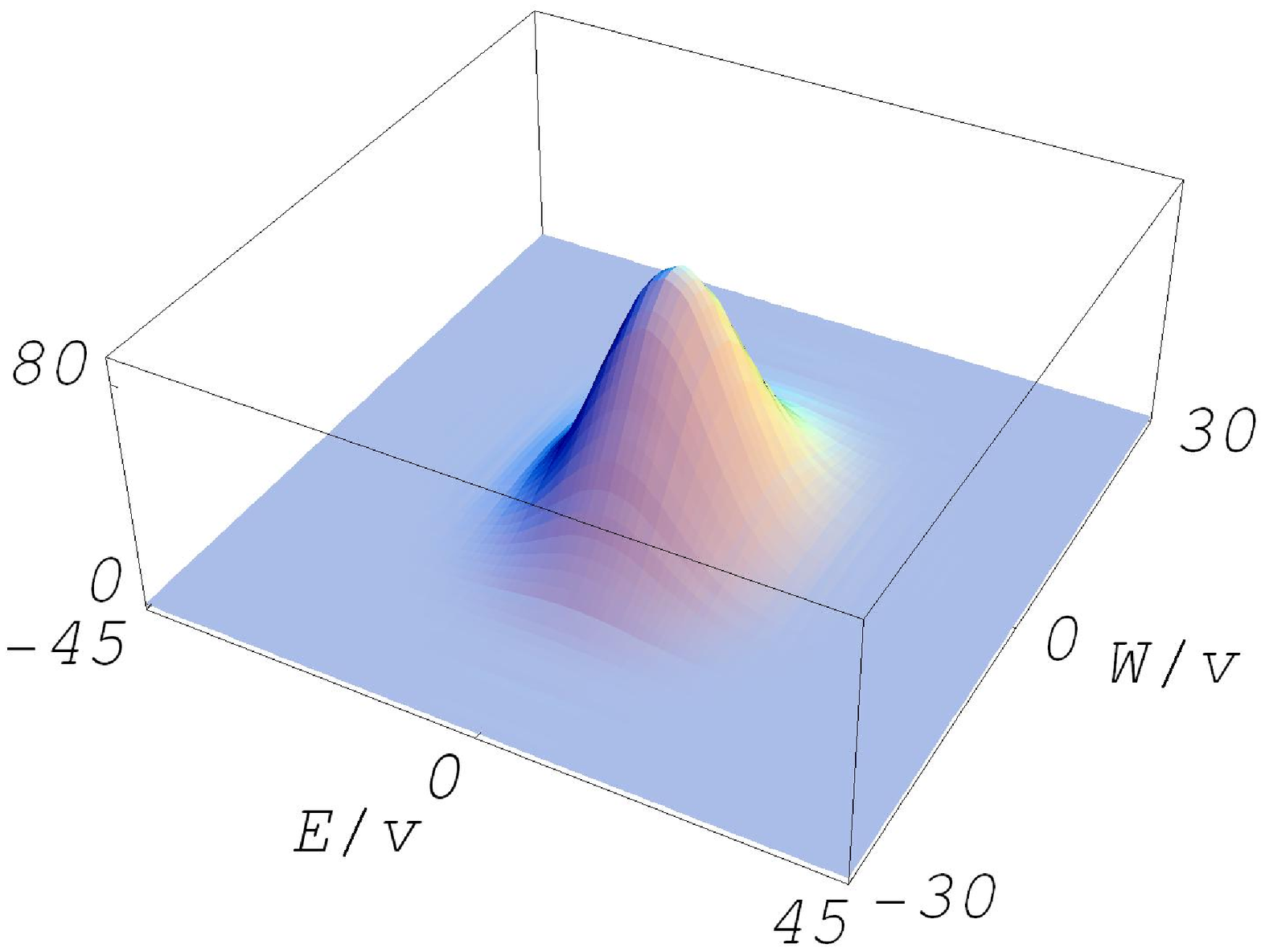}
\includegraphics[width=0.3\textwidth]{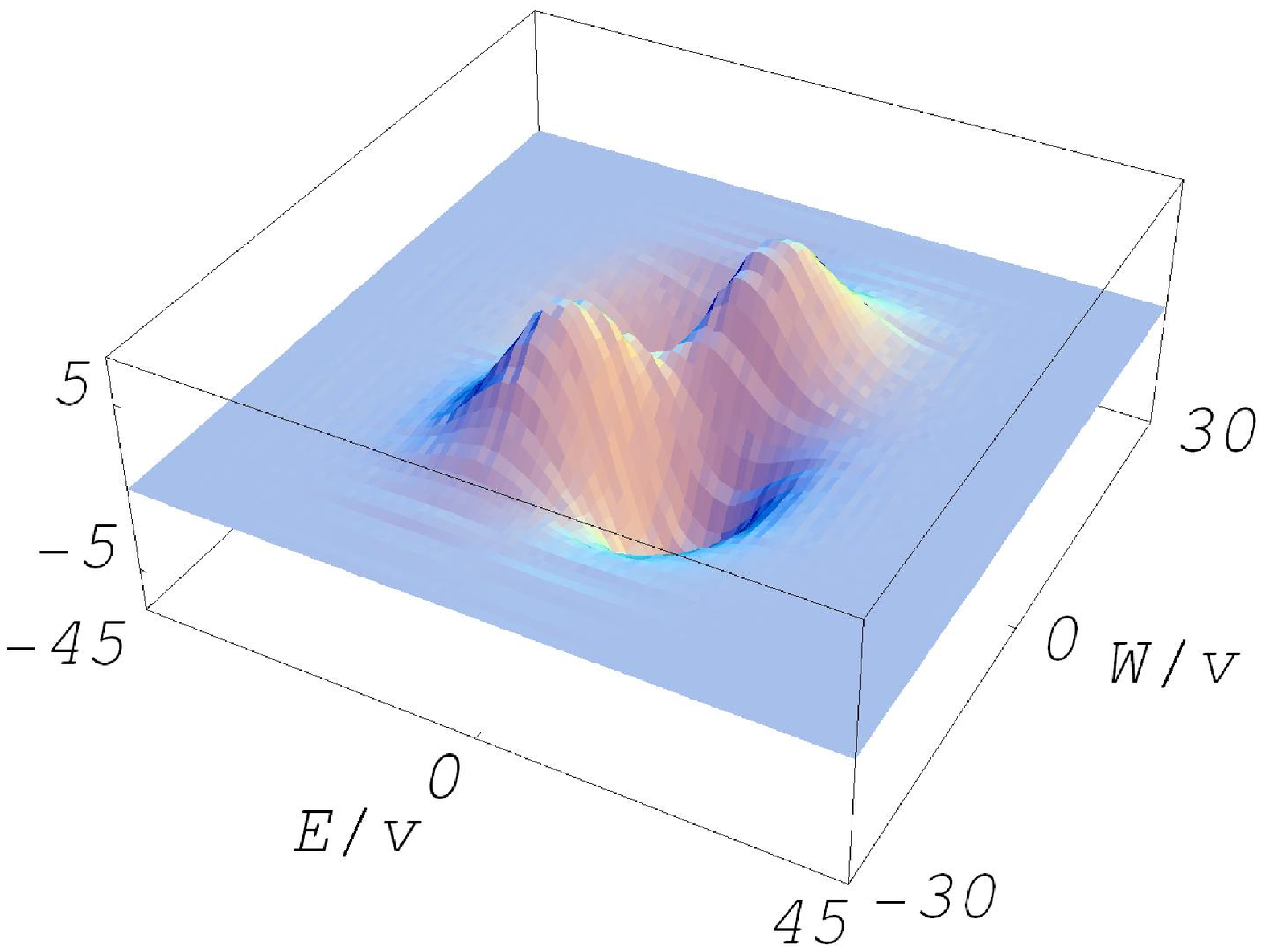}
\caption{\label{fig12} Same as Fig.~\ref{fig11} but for the $J=0,
T=0$ states of $^{24}$Mg and of $^{22}$Ne (from 400 realizations of
the ensemble).}
\end{figure*}

\end{widetext}

In view of the results in Subsection~\ref{info} and in Appendix~3, the
existence of density--density correlations is expected. Indeed, the
form~(\ref{7c}) of the TBRE Hamiltonian shows that the centroids of the
spectra must show particularly strong correlations. It is essentially
these correlations that are displayed in Figs.~\ref{fig11} and
\ref{fig12}. We ask: Are there also significant correlations between
the spectral fluctuations of levels with different spins and/or
different masses? These would show up in correlations between level
spacings and would, thus, be independent of the fluctuations of the
centroids of the spectra.  To answer this question, a finer test than
displayed in Figs. \ref{fig11} and \ref{fig12} is required.
We have not explored this question any further. We recall, however,
that data relevant to this question were published in Ref.~\cite{JBD}:
With $E_J$ the energy of the lowest state with spin $J$, the ratio
$\rho = (E_4 -E_2)/(E_2 -E_0)$ shows a broad peak in the range $\rho =
0$ to $\rho \approx 1$. This suggests the existence of level
correlations which go beyond those due to the centroids.

As remarked above, spectral fluctuations of spectra with different
quantum numbers go beyond the assumptions of canonical random--matrix
theory.  Therefore, it would be of interest to verify the existence of
such correlations experimentally. We have explored the possibility of
such a test in the following way. In nuclei the residual interaction
is fixed and not random. Therefore, ensemble averages cannot be taken.
In canonical random--matrix theory, one circumvents this problem using
ergodicity: The running average of an observable over a single
realization of a spectrum is equal to its ensemble average. In the
TBRE, we propose to take the average over an ensemble of nuclei in the
same shell as a substitute. In the spirit of this proposal (which
unfortunately lacks a theoretical foundation) we have calculated the
spacings between the lowest levels for a number of $sd$--shell nuclei
using the Brown-Wildenthal two--body interaction \cite{Bro88}.  The
ensemble consists of the nuclei $^{20-24}$Ne, $^{22-24}$Na,
$^{24-26}$Mg, $^{26}$Al, $^{30}$Si, $^{34}$P, $^{32,34}$S, and
$^{36}$Ar. For the even (odd) mass nuclei of this ensemble we
considered the correlations of the $J=0$ and $J=2$ ($J=1/2$ and
$J=5/2$) states. We label the
nearest-neighbor spacings of levels with equal spins consecutively by
$\Delta E_i$, $i = 1,2,3,...$, starting from the ground state.
Figure~\ref{fig14} shows, from left to right, the
average of the spacings $\overline{\Delta E_i \Delta E_j}$, the
product of the averages $\overline{\Delta E_i}\,\overline{\Delta
E_j}$, and the correlator $\overline{\Delta E_i \Delta E_j} -
\overline{ \Delta E_i}\,\overline{\Delta E_j}$, in units of
(MeV)$^2$. We see that correlations exist that are about 10\% of the
product of the averages.

\begin{widetext}

\begin{figure*}[h]
\includegraphics[width=0.3\textwidth]{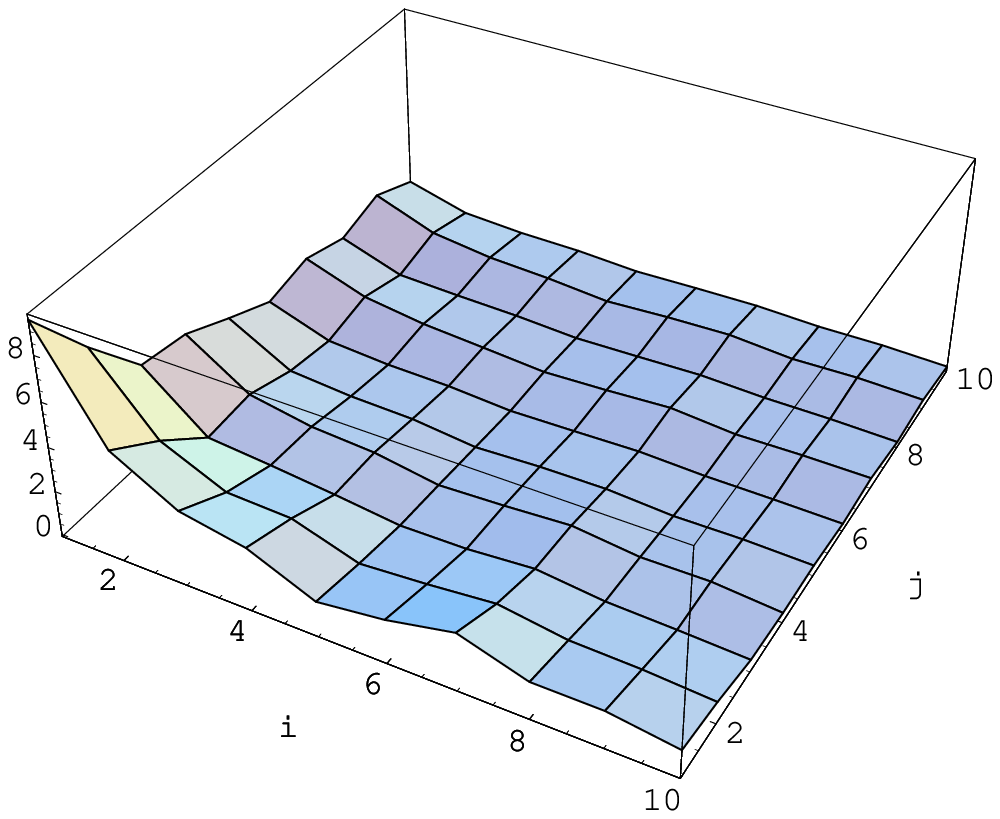}
\includegraphics[width=0.3\textwidth]{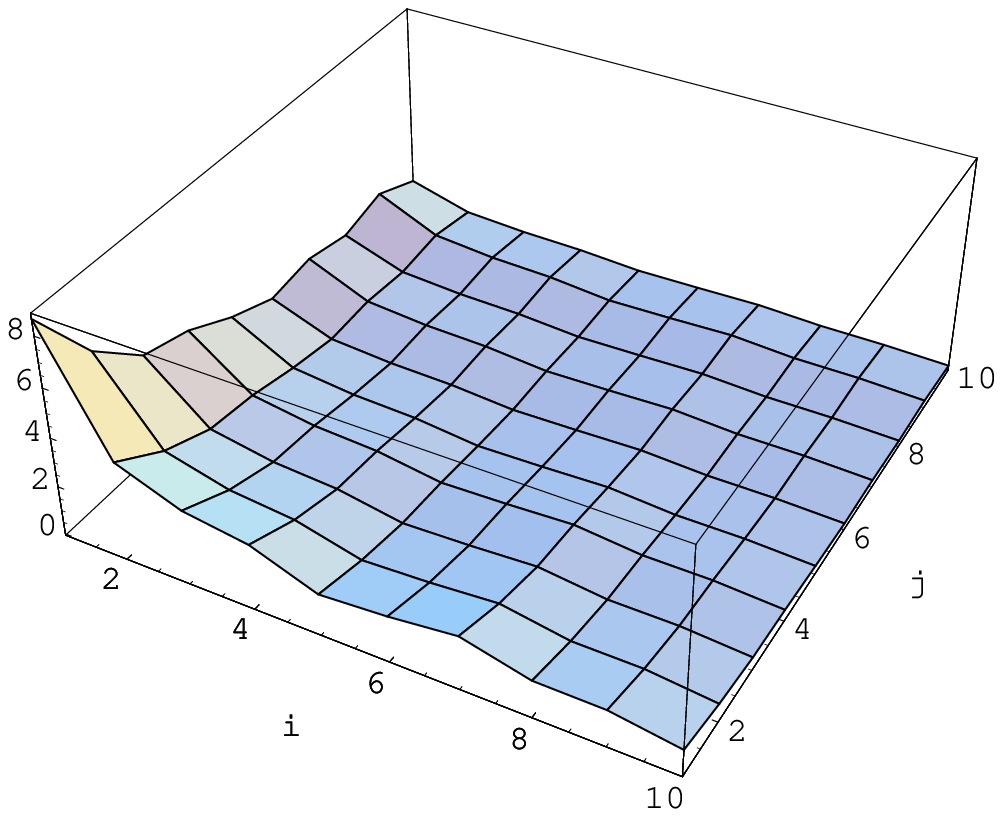}
\includegraphics[width=0.3\textwidth]{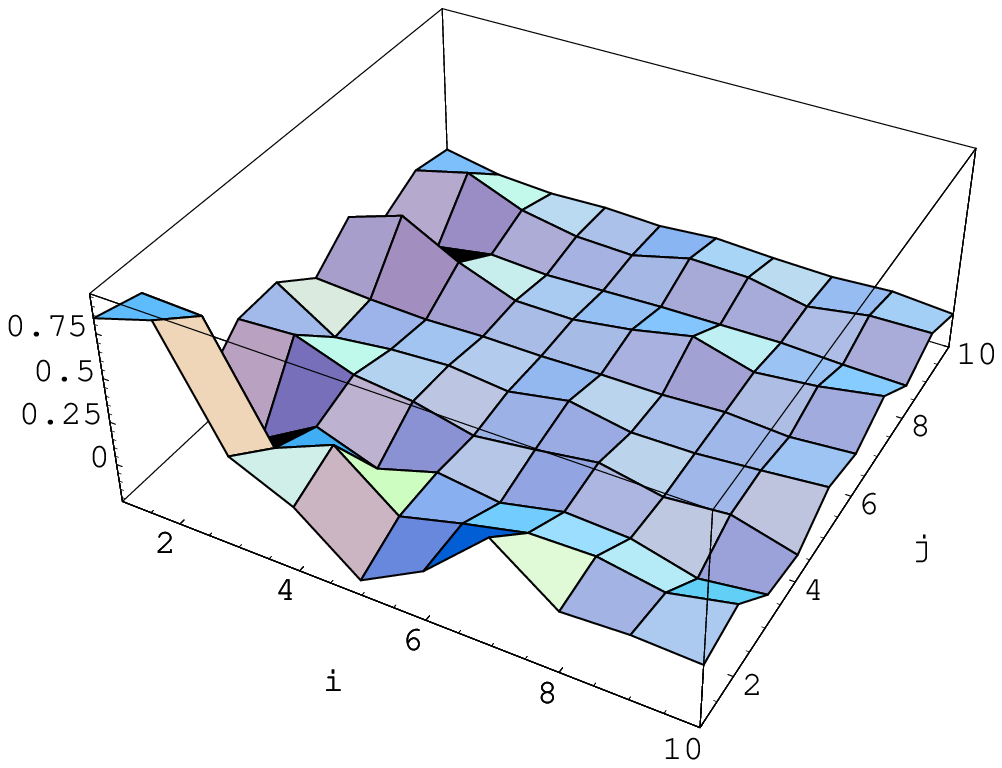}
\caption{\label{fig14} Level spacing correlations for 17 $sd$-shell
nuclei between spins $J=0$ and $J=2$ ($J=1/2$ and $J=5/2$ )for
even-$A$ (odd-$A$) nuclei. Left: Average of product of
spacings. Middle: Product of average spacings; Right: Correlation.}
\end{figure*}

\end{widetext}

The results of this subsection show that small correlations (of the
order of 10\%) exist between the spectral fluctuations of $sd$-shell
nuclei. We expect that the experimental verification of the existence
of these correlations is somewhat challenging, as it requires the
measurement of several complete spectral sequences involving 5-10
levels each.

\section{Summary}
\label{sum}

We have studied various aspects of the two--body random ensemble in
nuclei, and presented three main results. First, the geometric
properties of the nuclear shell model become transparent once one
replaces the usual two--body operators by linear combinations that
are orthogonal under the trace. This transformation yields one
operator that is approximately proportional to the unit matrix and
has the dominant spectral width. This monopole operator sets the
scale for nuclear binding, and can be derived analytically. The
remaining linear combinations of two--body operators have approximately
zero centroid and smaller spectral widths. They determine the spectral
fluctuation properties. Some of these operators have very small or even
zero widths and therefore are close to or equal to the null operator.
It is difficult or impossible to determine the corresponding linear
combinations of two--body matrix elements by fit to experimental data.
Thus, the geometric properties of the shell model explain the
difficulties encountered when residual interactions are determined by
fit.

Second, we presented analytical and numerical results regarding the
fluctuations of and correlations between the spectral widths of
shell--model operators. The total spectral width can be linked to the
spectral radius through a simple scale factor. This approach allowed us
to give a semi--quantitative explanantion for the preponderance of
spin--zero ground states both for a single $j$--shell and for the 
$sd$--shell. 

Third, we studied correlations between spectral fluctuations of
spectra belonging to different spins and/or mass numbers. Our numerical
results for the $sd$--shell indicate that correlations exist at the
10\% level in the region of low--lying excitations. This is new
territory and goes beyond the assumptions of canonical random--matrix
theory. It would be of considerable interest to explore how these
correlations depend on matrix dimension and, thus, change as we
consider other major shells.

We have shown that the complex geometric structure of the shell model
can be understood in rather simple terms. Apart from their theoretical
interest, our results might be of practical use in fitting effective
interactions, and they may motivate an experimental test of the
correlations between spectra belonging to different quantum numbers.

\section*{Appendix~1. Two--body Interaction}
\label{app1}

For simplicity of notation, we confine ourselves to the case of a single
$j$--shell and identical nucleons, omitting isospin and parity quantum 
numbers. The generalization is straightforward.

Let $a^{\dagger}_{\mu}$ and $a^{}_{\mu}$ be the creation and annihilation
operators for a particle with half--integer total spin $j$ and
$z$--component $\mu$. Two such identical particles are coupled to total
spin $s$ with $s = 0,2,\ldots,2j - 1$,
\be
A^s_M = \sum_\mu c(j j s; \mu ,M - \mu) a^{}_\mu a^{}_{M - \mu} \ .
\label{A1}
\ee
This is a tensor of rank $s$ which is non--zero if $s$ is even. It
transforms under rotations with the complex conjugate Wigner
$D$--function,
\be
(A^s_M)_{new} = \sum_N (D^s_{M N})^* (A^s_N)_{old} \ .
\label{A2}
\ee
The adjoint operator is
\be
(A^s_M)^{\dagger} = - \sum_\mu c(j j s; \mu ,M - \mu) a^{\dagger}_\mu
a^{\dagger}_{M - \mu} \ .
\label{A3}
\ee
This operator transforms with the usual $D$--function,
\be
(A^s_M)_{new} = \sum_N (D^s_{M N}) (A^s_N)_{old} \ .
\label{A4}
\ee
From the unitarity of the $D$--functions, it follows that
\be
T(s) = \sum_M (A^s_M)^{\dagger} A^s_M
\label{A5}
\ee
is a scalar.

It is convenient to rearrange the order of the creation and annihilation
operators,
\begin{eqnarray}
T(s) &=& - \sum_{\mu M \nu} c(j j s; \mu, M - \mu) c(j j s; \nu, M - \nu)
\nonumber\\
&&\times a^{\dagger}_{M -\mu} a^{}_\nu a^{\dagger}_\mu a^{}_{M - \nu} 
\nonumber \\
&& + \sum_{\mu M} c^2(j j s; \mu, M - \mu) a^{\dagger}_{M -\mu}
a^{}_{M - \mu} \nonumber \\
&=& - \sum_{\mu M \nu} c(j j s; \mu, M - \mu) c(j j s; \nu, M - \nu)
\nonumber\\
&&\times a^{\dagger}_{M -\mu} a^{}_\nu a^{\dagger}_\mu a^{}_{M - \nu} 
\nonumber \\
&& + \frac{2j + 1}{2s + 1} \sum_{\mu} a^{\dagger}_{\mu} a^{}_{\mu}
\ .  
\label{A6}
\end{eqnarray}
With $m$ the number of particles in the shell, the last term is equal
to $m (2j + 1)/(2s + 1)$. This term is denoted by $A(s)$ and gives a
diagonal contribution to the Hamiltonian. We define the tensor operators
${\cal A}^t$ of rank $t$
\begin{equation}
{\cal A}^t_M = \sum_{\rho} c(j j t; \rho, M - \rho) a^{\dagger}_\rho
a^{}_{\rho - M} \ .
\label{A7}
\end{equation}
For $t = 0$, the tensor operator ${\cal A}^t$ is just the number
operator, for $t = 1$, it is the operator of total spin, etc. Using
standard transformations of the Clebsch--Gordan coefficients, we
rewrite the terms in Eq.~(\ref{A2}) as 
\ba
T(s) &=& (2 s + 1 ) \sum_t W(j j j j; s t) \sum_{M} (-)^{s + 2j
  + M} {\cal A}^t_M {\cal A}^t_{- M} \nonumber\\
&&+ A(s) \ .
\label{A8}
\ea
With these definitions and with $\alpha = 1 + s/2$, the two--body
interaction $V_2$ has the form
\be
V_2 = \sum_{\alpha = 1}^{j+1/2} v_\alpha T(\alpha)
\label{A9}
\ee
where we use either the definition~(\ref{A5}) or the
definition~(\ref{A8}) for $T(s)$.

\section*{Appendix~2. Some properties of the matrices $C_{\mu
\nu}({\bf J})$}
\label{app2}

As in Appendix~1, we confine ourselves to the case of a single
$j$--shell and identical nucleons, omitting isospin and parity quantum 
numbers. We use the definition~(\ref{A1}) and (\ref{A3}).

To work out the properties of the matrices $C_{\mu \nu}({\bf J})$,
we shall presently need the commutator $[T(s), T(t)]$ for $s \neq t$,
with $T(s)$ defined in Eq.~(\ref{A5}). That commutator is given by
\ba
[T(s), T(t)] &=& \sum_{M,N} \bigg((A^s_M)^{\dagger} [A^s_M ,
(A^t_N)^{\dagger}] A^t_N \nonumber \\ && \qquad + (A^t_N)^{\dagger}
[(A^s_M)^{\dagger}, A^t_N] A^s_N \bigg) \ .    
\label{B1}
\ea
The commutator of two $A$'s can be expressed in terms of the
irreducible tensor operators ${\cal A}$ introduced in Eq.~(\ref{A7}).
We find
\ba
[(A^s_M)^{\dagger}, A^t_N] &=& -4 \sum_p \sqrt{(2s+1)(2p+1)} W(s j
t j; j p) \nonumber\\
&& \times c(s t p; M, N) B^p_{M - N} \ . 
\label{B2}
\ea
Here $B^p_m$ is an irreducible tensor of rank $p$ defined as
\be
B^p_m = \sum_n (-)^{j - n} c(j j p; -n, n - m) a^{\dagger}_n
a^{}_{n - m} \ . 
\label{B3}
\ee 
Hence,
\ba
[T(s), T(t)] &=& - \sum_{p} \sqrt{\frac{(2s+1)(2p+1)}{2t+1}}
W(s j t j; j p) \nonumber \\
&& \qquad \times \sum_{M,N} c(s t p; M, N) (A^t_N)^{\dagger} B^p_{M
- N} A^s_M \nonumber \\
&& + \sum_p \sqrt{\frac{(2t+1)(2p+1)}{2s+1}} W(t j s j; j p)
\nonumber \\
&& \qquad \times \sum_{M,N} c(t s p; N, M) (A^s_M)^{\dagger} B^p_{N
- M} A^t_M \ .\nonumber\\
\label{B4}
\ea
The result is a scalar because the three operators $A^{\dagger}, B$
and $A$ are coupled to rank zero. It is obvious that the commutator
$[T(s), T(t)]$ differs from zero and does not lie in the linear space
spanned by the operators $T$. A recoupling of the operators shows,
in fact, that the commutator is a sum of three--body operators.

The elements of the matrices $C_{\mu \nu}(J, s)$ with $s = 2\alpha -
2$, may be viewed as matrix elements of operators ${\cal C}(J,s)$.
The latter are defined in the space of Slater determinants of
single--particle shell--model states (i.e., states not coupled to
total spin $J$) and have the form
\be
{\cal C}(J,s) = {\cal P}(J) T(s) {\cal P}(J)
\label{B5}
\ee
where ${\cal P}(J)$ is the orthonormal projector onto the subspace of
many--body states with fixed total spin $J$. The projectors obey
\be
{\cal P}(J_1) {\cal P}(J_2) = {\cal P}(J_1) \delta_{J_1 J_2} \ ; \
[{\cal P}(J)]^{\dagger} = {\cal P}(J) \ . 
\label{B6}
\ee
With the help of the operator ${\vec J}$ of total spin, ${\cal P}(J)$
is explicitly written as
\be
{\cal P}(J) = \prod_{J_l \neq J} \frac{{\vec J}^2 - J_l (J_l + 1)}{J
(J + 1) - J_l (J_l + 1)} \ .
\label{B7}
\ee
The multiplication extends over all possible values $J_l$ of total
spin.

It is straightforward to show that the operators ${\cal P}(J)$ commute
with $T(s)$. Therefore, we have
\be
{\cal C}(J,s) = {\cal P}(J) T(s) {\cal P}(J) =  T(s) {\cal P}(J) =
{\cal P}(J) T(s) \ .
\label{B8}
\ee
It follows that the commutator $[ {\cal C}(J,s), {\cal C}(J,t)]$ can
be written as
\be
[ {\cal C}(J,s), {\cal C}(J,t)] = {\cal P}(J) [T(s), T(t)] \ .
\label{B9}
\ee
But we have shown in Eq.~(\ref{B4}) that the commutator $[T(s), T(t)]$
does not vanish and is, in fact, a sum of three--body operators. It
then follows from Eq.~(\ref{B9}) that the operators ${\cal C}$ do not
commute. Rather, their commutators represent sums of three--body
operators (projected, of course, onto the space with fixed $J$). In
particular, it is not possible to diagonalize the ${\cal C}$'s
simultaneously.

\section*{Appendix~3. Another Form of the TBRE}

As mentioned in Subsection~\ref{info}, there exists another possibly
useful form of the TBRE. This we now derive. For $\alpha = 1, \ldots,
a$ we define
\be
c({\bf J}, \alpha) = {\rm Trace} \ C_{\mu \nu}({\bf J}, \alpha)
\label{7a}
\ee
and introduce the traceless matrices
\be
{\tilde C}_{\mu \nu}({\bf J}, \alpha) = C_{\mu \nu}({\bf J}, \alpha)
- \frac{c({\bf J}, \alpha)}{d({\bf J})} \delta_{\mu \nu} \ .
\label{7b}
\ee
With these definitions, the Hamiltonian in Eq.~(\ref{1}) takes the
form
\be
H_{\mu \nu}({\bf J}) = {\tilde H}_{\mu \nu}({\bf J}) + \bigg(
\sum_{\alpha = 1}^a \frac{c({\bf J}, \alpha)}{d({\bf J})} v_\alpha
\bigg) \delta_{\mu \nu}
\label{7c}
\ee
where
\be
{\tilde H}_{\mu \nu}({\bf J}) = \sum_\alpha v_\alpha {\tilde C}_{\mu
\nu}({\bf J}, \alpha)
\label{7d}
\ee
is traceless. We can now diagonalize the real symmetric positive
semidefinite $a$--dimensional matrix
\be
{\tilde S}_{\alpha \beta} = \frac{1}{d({\bf J})} {\rm Trace} \ [
{\tilde C}({\bf J}, \alpha) {\tilde C}({\bf J}, \beta) ]
\label{7e}
\ee
by an orthogonal transformation ${\tilde {\cal O}}_{\alpha \beta}
({\bf J})$. Of the resulting eigenvalues ${\tilde s}^2({\bf J},
\alpha)$, exactly $(a_1 - 1)$ differ from zero. Using the same
construction as in Eqs.~(\ref{3}) to (\ref{5}) and denoting the
resulting $a_1 - 1$ orthonormal matrices by ${\tilde B}({\bf J},
\alpha)$, we arrive at
\ba
H_{\mu \nu}({\bf J}) &=& \sum_{\alpha = 1}^{a_1 - 1}
{\tilde w}_\alpha({\bf J}) {\tilde s}_\alpha({\bf J}) {\tilde B}({\bf
J}, \alpha) \nonumber\\
&+& \delta_{\mu \nu} \sum_{\alpha = 1}^a \frac{c({\bf J},
\alpha)}{d({\bf J})} v_\alpha \ .
\label{7f}
\ea
Similarly as in Eq.~(\ref{7}) but with ${\cal O}$ replaced by
$\tilde{\cal O}$, the $(a_1 - 1)$ random variables
${\tilde w}_\alpha({\bf J})$ are linear combinations of the
$v_\alpha$'s. They are uncorrelated, have mean values zero, and a
common second moment equal to unity. The matrices ${\tilde B}({\bf
J}, \alpha)$ are traceless and, hence, orthogonal to the unit matrix
in the sense of the trace norm. Thus, Eq.~(\ref{7f}) is another
representation of the TBRE in terms of $a_1$ orthonormal matrices.
In contrast to Eq.~(\ref{6}), we have now separated a term
(proportional to the unit matrix) which defines the centroid of the
spectrum but has no bearing on the actual level sequence, from the
remainder which in turn possesses a vanishing centroid but yields
the actual details of the spectrum. The form~(\ref{7f}) is probably
not useful for data analyses but should be helpful whenever we are
interested in spectral fluctuation properties. These depend only
upon the $a_1 - 1$ variables ${\tilde w}_\alpha({\bf J})$.

\section*{Appendix~4. Approximate Diagonalization of $S_{\alpha
\beta}$}
\label{salpha}

As in previous Appendices, we perform the calculation for a single
$j$--shell with identical nucleons. The number of nucleons is denoted
by $m$. We use the fact that here we have $s = 2\alpha -2$ and replace
$\alpha, \beta$ by $s, t$. We use the definition~(\ref{B8}) and the
fact that $P(J)$ and $T(s)$ commute to write $S_{s t}$ in the form
\be
S_{s t}(J) = \frac{1}{d(J)} {\rm Trace} \ [ {\cal P}(J) T(s) T(t) ] \ .
\label{C1}
\ee
The calculation is difficult for fixed total spin $J$. However, our
numerical calculations show that the eigenvalues of $S_{s t}(J)$ and
the eigenvector pertaining to the largest eigenvalue change little
with $J$ for the lowest values of $J$. Moreover, these lowest values
of $J$ have the largest values of $d(J)$. Therefore, we approximate
the calculation of $S_{s t}(J)$ by averaging over all values of $J$,
thus defining
\be
X_{s t} = \frac{1}{\sum_J d(J)} \sum_J d(J) S_{s t}(J) \ .
\label{C2}
\ee
We observe that the sum over $J$ removes the projectors ${\cal P}(J)$
since their sum equals unity. Thus, the average is equal to the
normalized trace of $T(s T(t)$ taken with respect to all states in
the Hilbert space spanned by normalized Slater determinants (without
any attention paid to the coupling of spins). We denote that trace by
angular brackets and have
\be
X_{s t} = \langle T(s) T(t) \rangle \ .
\label{C2a}
\ee
We calculate $X_{s t}$ approximately by keeping only the leading terms
in an expansion in inverse powers of $(2j+1)$. That is the meaning of
the approximate sign in this Appendix.

The trace in Eq.~(\ref{C2a}) can be worked out using Wick contraction.
The relevant term is
\be
\langle a^{\dagger}_{\mu} a^{\dagger}_{M - \mu} a^{}_{\nu} a^{}_{M
- \nu} a^{\dagger}_{\rho} a^{\dagger}_{N - \rho} a^{}_{\sigma} a^{}_{N
- \sigma} \rangle \ .
\label{C3}
\ee
Two types of Wick contractions occur: (a) Those where contractions
connect pairs of operators separately within the group of the first
four and of the second four operators, and (b) those where at least two
operators of the first group of four are contracted with the same
number of operators of the second group. Among all these, the terms of
leading order in an expansion in inverse powers of $(2j+1)$ are the
terms with the maximum number of unconstrained summations over magnetic
quantum numbers. Every Wick contraction of a pair of operators produces
a Kronecker delta. However, the two Wick contractions in case (a)
affecting the first (the second) group of four operators produce the
same Kronecker delta's $\delta_{\mu \nu}$ ($\delta_{\rho \sigma}$,
respectively). This is not so in case (b) where each of the two Kronecker
delta's imposes a new constraint. We conclude that the terms of leading
order arise only from Wick contractions for case (a). This fact implies
that
\be
X_{s t} \approx \langle T(s) \rangle \ \langle T(t) \rangle = f(s) f(t)
\label{C3a}
\ee
where we have defined
\be
f(s) = \langle T(s) \rangle \ .
\label{C3b}
\ee
Factorization of $X$ as in Eq.~(\ref{C3a}) implies that the leading
eigenvalue $s^2_1$ defined in Eq.~(\ref{3}) is given by
\be
s^2_1 \approx \sum_s f^2(s)
\label{C3c}
\ee
while all other eigenvalues are $\approx 0$. The associated matrix $B_1$
is the matrix representation of an operator ${\cal B}_1$. The latter is
proportional to $\sum_s f(s) T(s)$. Normalization requires $\langle B^2_1
\rangle = 1$. We use again that the Wick contractions approximately
factorize and find that
\be
{\cal B}_1 \approx \frac{1}{\sum_s f^2(s)} \ \sum_s f(s) T(s) \ .
\label{C3d}
\ee
We observe that $\langle B_1 \rangle \approx 1$. The same statement
holds, within our factorization approximation, for all higher powers
of $B_1$. Thus, $B_1$ is (except for normalization) approximately
equal to the unit matrix. With the dimension of total Hilbert space
given by $N(j,m) = {2j+1 \choose m}$, this gives
\be
B_1 \approx N^{-1/2}(j,m) \ {\bf 1} \ .
\label{C3e}
\ee

It remains to work out the function $f(s)$. We denote the Wick
contraction by an index $W$. We have
\be
\bigg( a^{\dagger}_{\mu} a^{\dagger}_{M - \mu} a^{}_{\nu} a^{}_{M -
  \nu} \bigg)_W =  [ \delta_{\mu \nu} - \delta_{\mu M - \nu} ]
  a^{\dagger}_{\mu} a^{\dagger}_{M - \mu} a^{}_{\mu} a^{}_{M - \mu} \ .
\label{C4}
\ee
Thus,
\ba
f(s) &=& \langle T(s) \rangle \approx \sum_{M \mu \nu} c(jjs;\mu,M-\mu)
c(jjs;\nu,M-\nu) \nonumber\\
&&\times [ \delta_{\mu \nu} - \delta_{\mu M - \nu} ] \bigg(
\frac{m}{2j+1} \bigg)^2
\label{C4a}
\ea
or
\be
f(s) \approx \bigg( \frac{m}{2j+1} \bigg)^2 \ 2 \ (2s+1) \ .
\label{C4b}
\ee
For $j = 19/2$ and $m = 6$ ($m=8$), this yields for $s^2_1$ the value
$39.933$ ($126.21$) so that $s_1 = 6.32$ ($11.24$, respectively).

With $f(s)$ given by Eq.~(\ref{C4b}), we can work out ${\cal B}_1$
explicitly using Eq.~(\ref{C3d}) and Eq.~(\ref{A6}) for $T(s)$,
omitting normalization factors. This yields
\ba
{\cal B}_1 &\propto& \frac{1}{(\sum_s (st+1)^2)^{1/2}} \bigg\{
\frac{1}{2} (2j+1)^2 m \ {\bf 1} \nonumber \\
&& - \sum_{s M \mu \nu} (2s+1) c(jjs; \mu, M - \mu)
c(jjs; \nu, M - \nu) \nonumber\\
&&\times a^{\dagger}_{M - \mu} a^{}_\nu a^{\dagger}_\mu
a^{}_{M - \nu} \bigg\} \ . 
\label{C5}
\ea
We retrieve the unit matrix as in Eq.~(\ref{C3e}) and conclude that
compared to the laeding term, the last sum in Eq.~(\ref{C5}) is of
order $1 / (2j+1)$.

These results allow us to estimate $s^2_1(J)$. We do so by assuming
that the results obtained by averaging over all values of $J$ apply
separately to each $J$. Since $B(J, 1)$ is, except for terms of order
$1/(2j+1)$, equal to a multiple of the unit matrix, we equate the
variance of the coefficient multiplying $B(J, 1)$ in Eq.~(\ref{6})
with the variance of the coefficient multiplying $d^{-1/2}(J)
\delta_{\mu \nu}$ in Eq.~(\ref{7f}). We obtain
\be
s^2_1(J) \approx (1 / d(J)) \sum_\alpha c^2(J, \alpha) \ .
\label{7g}
\ee
By the same token, we expect that all the $a_1-1$ eigenvalues
${\tilde s}^2_\alpha(J)$ in Eq.~(\ref{7e}) are on average of equal
size and of order $1/(2j+1)$. This implies that the statistical
fluctuations of the centroid of the spectrum are considerably larger
than those of individual level spacings.

In summary, we have shown that the leading eigenvalue and
eigenfunction of $X_{s t}$ are easily obtained because for $2j+1 \gg
1$, $X_{s t}$ approximately factorizes. The factors $f(s)$ and
$f(t)$ are simply given by Wick--contracting the operators $T(s)$ and
$T(t)$ separately. The leading eigenvalue is given by Eq.~(\ref{C3c})
and the matrix $B_1$ is approximately given by the unit matrix.

This research was supported in part by the U.S. Department of Energy
under Contract Nos.\ DE-FG02-96ER40963 (University of Tennessee) and
DE-AC05-00OR22725 with UT-Battelle, LLC (Oak Ridge National
Laboratory).

\end{document}